\documentclass[useAMS,usenatbib,twocolumn]{mn2e}
\usepackage[]{graphicx}
\usepackage{latexsym}
\usepackage{amssymb}
\usepackage{float}
\usepackage{ifthen}
\usepackage{url}

\newcommand {\hI} {$\mathrm{H_{I}}$}
\newcommand {\hII} {$\mathrm{H_{II}}$}
\newcommand {\nII} {$\mathrm{N_{II}}$}
\newcommand {\ha} {H$\alpha$}
\newcommand {\has} {H$\alpha$}
\newcommand {\kms} {\,km\,s$^{-1}$\,}

\newcommand{\Min}{${}^{\prime}$}
\newcommand{\Sec}{${}^{\prime\prime}$}
\newcommand{\Deg}{${}^{\circ}$}

\newcommand{\Arc}{''}
\newcommand{\PA}{$\mathrm{PA}$}
\def\degr{\hbox{$^\circ$}}
\newcommand{\BB} {\textit{\texttt{B}\texttt{\has}\texttt{BAR}}}
\newcommand{\FM} {\texttt{\textsc{FaNTOmM}}}
\newcommand{\um} {$\mu$m\,}
\newcommand{\NN} {28}

\title[\ha\, Kinematics of the SINGS Nearby Galaxies Survey. I]{\ha\, Kinematics of the SINGS Nearby Galaxies Survey. I\thanks{Based on observations collected at the European Southern Observatory, La Silla, Chile.}}
\author[O. Daigle et al.]{O. Daigle$^{1,5}$\thanks{E-mail:
odaigle@astro.umontreal.ca}, C. Carignan$^{1,5}$, P. Amram$^{2,5}$, 
O. Hernandez$^{1,2,5}$, L. Chemin$^{1}$,
\newauthor C. Balkowski$^{3,5}$ and R. Kennicutt$^{4}$\\
$^{1}$Observatoire du mont M\'egantic, LAE, Universit\'e de Montr\'eal, 
C. P. 6128 succ. centre ville, Montr\'eal, Qu\'ebec, Canada H3C 3J7.\\
$^{2}$Observatoire Astronomique de Marseille Provence, Laboratoire d'Astrophysique de Marseille, 2 place 
Le Verrier, F-13248 Marseille\\
Cedex 04, France.\\
$^{3}$Observatoire de Paris, section Meudon, GEPI, CNRS UMR 8111 et Universit\'e 
Paris 7, 5 place J. Janssen, 92195 Meudon Cedex,\\France.\\
$^{4}$Department of Astronomy, Steward Observatory, 933 N. Cherry Ave., Tucson, AZ 85721-0065, USA.\\
$^{5}$Visiting Astronomer, Canada--France--Hawaii Telescope, operated by the National Research Council of Canada, the Centre National\\de la Recherche Scientifique de France, and the University of Hawaii.}

\begin{document}

\date{Accepted . Received ; in original form }

\pagerange{\pageref{firstpage}--\pageref{lastpage}} \pubyear{2005}

\maketitle

\label{firstpage}

\begin{abstract}

This is the first part of an \ha\, kinematics follow-up survey of the SINGS sample. The data
for \NN\, galaxies are presented. The observations 
were done on three different telescopes with \FM, an integral field photon counting spectrometer,
installed in the respective focal reducer of each telescope. The data reduction was done through
a newly built pipeline with the aim of producing the most homogenous data set possible.
Adaptive spatial binning was applied to the data cubes in order to get a constant 
signal-to-noise ratio across the field of view. Radial 
velocity and monochromatic maps were generated using a new algorithm and the kinematical 
parameters were derived using tilted-ring models.

\end{abstract}


\section{Introduction}

The Legacy survey SINGS (Spitzer Infrared Nearby Galaxies Survey) wants to characterise the infrared 
emission across the entire range of galaxy properties and star formation environments, 
including regions that until now have been inaccessible at infrared wavelengths 
(\citeauthor{2003PASP..115..928K}, \citeyear{2003PASP..115..928K}). SINGS will provide: 
\begin{itemize}
\item new insights into the physical processes connecting star formation to the ISM properties 
of galaxies;
\item a vital foundation of data, diagnostic tools, and astrophysical inputs for understanding 
SPITZER observations of the distant universe and ultraluminous and active galaxies;
\item an archive that integrates visible/UV and IR/submillimeter studies into a coherent 
self-consistent whole, and enables many follow-up investigations of star formation and of the ISM.
\end{itemize}

The SPITZER observations will provide images in 7 different bands from 3.6\um to 160\um and 
spectroscopic data at medium and low resolution in the range 5--95\um. These data will be 
used to trace the distribution and content of different dust components, from the PAHs and 
very small grains in the mid-IR, to the big grains in the far-IR (\citeauthor{2003MNRAS.346..403D} \citeyear{2003MNRAS.346..403D}, \citeauthor{2003PASP..115..928K} \citeyear{2003PASP..115..928K}). Ancillary multiwavelength 
observations will provide images in X-rays, UV (1300--2800 \AA\ imaging and spectrophotometry), BVRIJHK, \ha, Pa$\alpha$, FIR, submillimeter, CO and \hI. A total of 20 ground- and space-based telescopes are providing supporting data.

These data will help understand the process of star formation and feedback mechanisms that 
are fundamental parameters regulating the formation and evolution of galaxies. History of 
star formation has been strongly different for galaxies of different morphological type 
and luminosity. While short events of star 
formation, probably triggered by violent merging, formed most stars in elliptical 
galaxies, late-type systems seem to have their star formation modulated by the angular 
momentum (\citeauthor{1986ARA&A..24..421S}, \citeyear{1986ARA&A..24..421S}) or by the mass of the initial system (\citeauthor{2001ApSSS.277..401B}, \citeyear{2001ApSSS.277..401B}). The 
process of star formation and feedback must thus be clearly understood in order to 
understand galaxies' evolution. However, these physical processes are still poorly 
known. The primordial atomic gas has to condense into molecular clouds to form stars. 
The newly formed stars inject metals into the interstellar medium via stellar winds, 
heat the dust and ionise the surrounding gas. It seems that the activity of star 
formation is regulated by the total gas surface density (\citeauthor{1989ApJ...344..685K}, \citeyear{1989ApJ...344..685K}), but it is 
still unclear what is the role of rotation in this process.

Even as important as it seems, no gathering of optical kinematical data was planned for the SINGS 
galaxies. This paper, by providing the \ha\, kinematics over the whole optical extent for \NN\, 
galaxies of the SINGS sample, wants to make up for this lack. A total of 58 SINGS galaxies are potentially observable in the \ha\, emission line (see section \ref{sample_section}). The \ha\, kinematics of the 30 remaining galaxies of the observable part of the sample will be published in a forthcoming paper. These data were obtained with \FM\, on three different telescopes (see section \ref{obs_runs_section}). \FM\, is an integral field spectrometer made of a photon-counting camera using 
a third generation photo-cathode, a scanning Fabry-Perot (FP) and a narrow-band interference filter. \FM\, was coupled to the focal reducer of the telescopes used. The photo-cathode used has a high quantum efficiency ($\sim$ 30\% at \ha). 
This camera enables one to scan rapidly ($\sim$ 5--10 minutes) the FP Free Spectral Range (FSR) and to cycle 
many times during an observation, thus averaging changing photometrical conditions, as compared 
to CCD observations where scanning must be done slowly to overcome the readout noise (details about the 
camera can be found in \citeauthor{2003SPIE.4841.1472H}, \citeyear{2003SPIE.4841.1472H} and \citeauthor{2002PASP..114.1043G}, \citeyear{2002PASP..114.1043G}). In this paper, section 2 gives an overview of the observational campaign and of the galaxies studied. Section 3 discusses 
how the data were reduced, processed and how the kinematical data and parameters were extracted. 
Section 4 provides all the maps extracted from the work done in section 3. Section 5 discusses the advantages of FP kinematical data as compared to other kinematical data. A short appendix is added to comment the observational characteristics of the galaxies presented. Once completed, the data set will be available in the SINGS database, as for the other SINGS ancillary surveys.


\begin{figure}
\begin{center}
\includegraphics[width=0.5\textwidth]{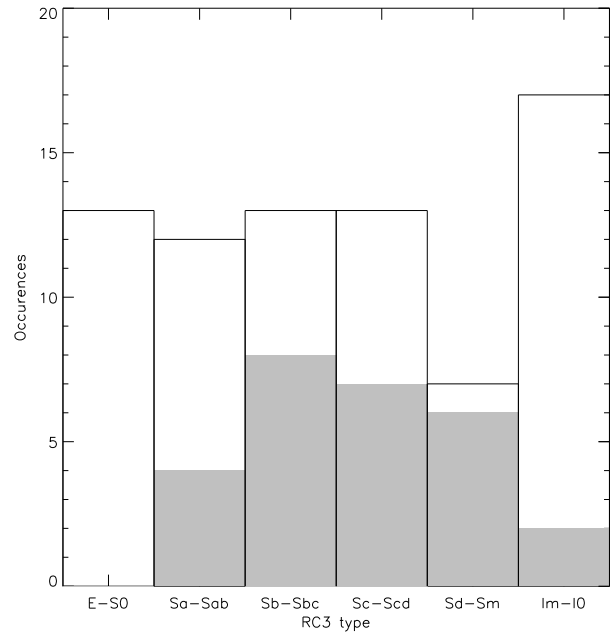}
\caption{The SINGS RC3 galaxy type distribution. The grey area shows the galaxies presented in this paper.}
\label{histo_figure}
\end{center}
\end{figure}

\begin{figure*}
\includegraphics[width=\textwidth]{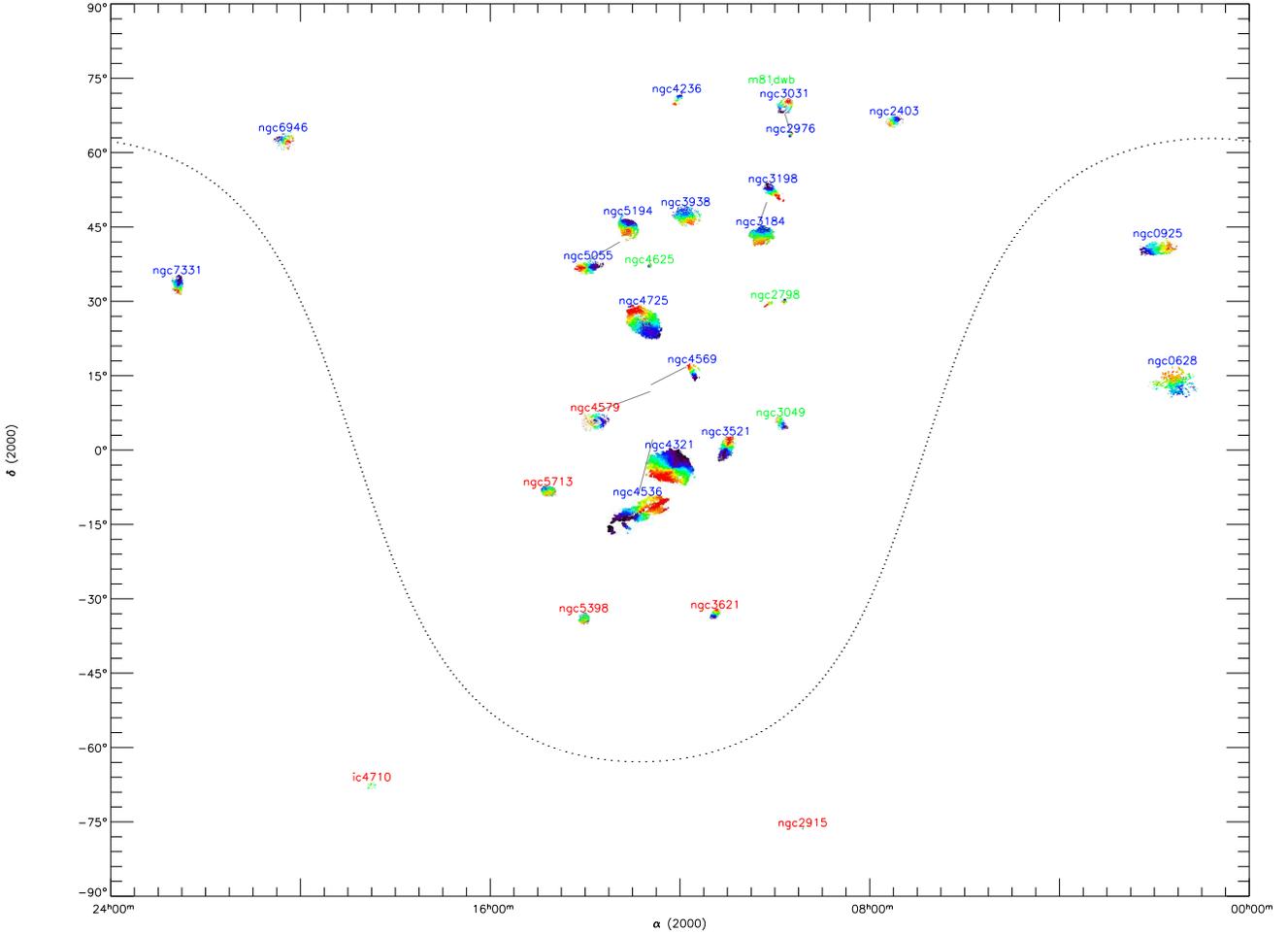}
\caption{The SINGS \ha\, kinematics sky coverage so far. The names in blue correspond to the 
observations done at the OMM, in green at the CFHT and in red at the ESO 3.6m
telescope. Galaxies  relative sizes are to scale. The galactic equator is shown as 
the dotted line. Naturally, there is a strong concentration toward the Virgo cluster.}
\label{great_map_figure}
\end{figure*}

\setcounter{table}{0}
\begin{table*}
\caption{Observational data for the SINGS \ha\, kinematics sample.} 
\begin{tabular}{ccccccccccc}
\hline \hline
Galaxy  &  $\alpha$(J2000)& $\delta$(J2000) & Type & $\Delta^{(1)}$ & $D_{25}^{b,i}$ $^{(2)}$ &  $B_T^{b,i}$ $^{(3)}$ &  $M_B^{b,i}$ $^{(4)}$ & Systemic Velocity$^{(5)}$\\
Name & (hh mm ss) & ($^o$ ' ") & RC3 & (Mpc) &  ($'$) &   &  & (\kms)  \\  
\hline
NGC 628   &   01 36 41.8   &   +15 47 00   &   SA(s)c   &   11.4   &  10.5 x 9.5  & 9.95            & $-$20.33      &     657\\
NGC 925   &   02 27 16.8   &   +33 34 41   &  SAB(s)d   &  9.3     &      11.2    & 10.6            & $-$19.24      &     554\\  
NGC 2403  &   07 36 54.5   &   +65 35 58   &  SAB(s)cd  &     4.2  & 21.4         & 8.5             & $-$19.61      &     132\\  
NGC 2798  &   09 17 22.9   &   +41 59 59   &  SB(s)a pec&   24.7   &   2.6 x 1    & 13.04           & $-$18.92      &    1726\\
NGC 2915  &   09 26 11.5   &   $-$76 37 35   &  I0        &   2.7    &   1.9 x 1    & 13.25           & $-$13.91      &     468\\
NGC 2976  &   09 47 15.4   &   +67 54 59   &   SAc pec  &    3.5   &   5.9 x 2.7  & 10.82           & $-$16.90      &       3\\
NGC 3049  &   09 54 49.6   &   +09 16 18   &  SB(rs)ab  &   19.6   &   2.2 x 1.4  & 13.04           & $-$18.42      &    1494\\
NGC 3031  &   09 55 33.2   &   +69 03 55   &   SA(s)ab  &    3.5   &  26.9 x 14.1 & 7.89            & $-$19.83      &     $-$34\\
UGC 5423  & 10 05 30.6   &   +70 21 52   &  Im        &    3.5   &   0.9 x 0.6    & 15.19           & $-$12.53      &     350\\
NGC 3184  &   10 18 17.0   &   +41 25 28   &  SAB(rs)cd &    8.6   &   7.4 x 6.9  & 10.36           & $-$19.31      &     592\\
NGC 3198  &   10 19 54.9   &   +45 33 09   &  SB(rs)c   &  14.5    &  7.8         & 11.1            & $-$19.70      &     660\\ 
NGC 3521  &   11 05 48.6   &   $-$00 02 09   &  SAB(rs)bc &    9     &   11 x 5.1   & 9.83            & $-$19.94      &     805\\
NGC 3621  &   11 18 16.3   &   $-$32 48 45   &  SA(s)d    &   6.2    &   12.3 x 7.1 & 10.28           & $-$18.68      &     727\\
NGC 3938  &   11 52 49.4   &   +44 07 15   &  SA(s)c    &   12.2   &  5.4 x 4.9   & 10.90           & $-$19.53      &     809\\
NGC 4236  &   12 16 42.1   &   +69 27 45   &  SB(s)dm   &  2.2     & 16.4         & 9.5             & $-$17.21      &       2\\ 
NGC 4321  &   12 22 55.2   &   +15 49 23   &  SAB(s)b   & 16.1     & 7.4          & 10.0            & $-$21.03      &    1590\\ 
NGC 4536  &   12 34 27.1   &   +02 11 16   &  SAB(rs)bc &   25     &   7.6 x 3.2  & 11.16           & $-$20.82      &    1808\\ 
NGC 4569  &   12 36 49.8   &   +13 09 46   &  SAB(rs)ab &   20     &   9.5 x 4.0  & 10.26           & $-$21.24      &    $-$235\\ 
NGC 4579  &   12 37 43.6   &   +11 49 05   &  SAB(rs)b  &   20     &   5.9 x 4.7  & 10.48           & $-$21.18      &    1591\\ 
NGC 4625  &   12 41 52.7   &   +41 16 25   &  SAB(rs)m pec&  9.5   &   2.2 x 1.9  & 12.92           & $-$16.97      &     609\\
NGC 4725  &   12 50 26.6   &   +25 30 06   &  SAB(r)ab pec& 17.1   &   10.7 x 7.6 & 10.11           & $-$21.05      &    1206\\
NGC 5055  &   13 15 49.3   &   +42 01 45   &  SA(rs)bc  &    8.2   &  12.6 x 7.2  & 9.31            & $-$20.26      &     504\\
NGC 5194  &   13 29 52.7   &   +47 11 43   &  SA(s)bc pec&   8.2   &  11.2 x 6.9  & 8.96            & $-$20.61      &     463\\
NGC 5398  &   14 01 21.5   &   $-$33 03 50   &  SB(rs)dm  &   15     &   2.8 x 1.7  & 12.78           & $-$18.10      &    1216\\
NGC 5713  &   14 40 11.5   &   $-$00 17 21   &  SAB(rs)bc pec & 26.6 &   2.8 x 2.5  & 11.84           & $-$20.28      &    1883\\
IC 4710   &   18 28 38.0   &   $-$66 58 56   &  SB(s)m    &   8.5    &   3.6 x 2.8  & 12.5            & $-$17.14      &     741\\
NGC 6946  &   20 34 52.0   &   +60 09 15   &  SAB(rs)cd & 5.5      &  14.9        & 7.92            & $-$20.78      &      46\\ 
NGC 7331  &   22 37 04.1   &   +34 24 56   &   SA(s)b   &   15.7   &  10.5 x 3.7  & 10.35           & $-$20.63      &     816\\
\hline \cr
\end{tabular}
\begin{flushleft}
$^{(1)}\Delta$ : distance in Mpc, flow-corrected for H$_{0}$=70 \kms$Mpc^{-1}$, as presented in \citeauthor{2003PASP..115..928K} (\citeyear{2003PASP..115..928K}).\\
$^{(2)}D_{25}^{b,i}$ : optical diameter at the 25 magnitude/arcsecond$^{2}$ in B,
corrected for the effects of projection and extinction. Taken from the RC3.\\
$^{(3)}B_T^{b,i}$ : corrected total apparent magnitude in B. Taken from the RC3.\\
$^{(4)}M_B^{b,i}$ : corrected total absolute magnitude in B. Calculated from $\Delta$ and $B_T^{b,i}$.\\
$^{(5)}$Systemic velocity : galaxy's systemic velocity. Taken from \citeauthor{2003PASP..115..928K} (\citeyear{2003PASP..115..928K}).\\
\end{flushleft}
\label{sample_table}
\end{table*}

\begin{center}
\begin{table*}
\caption{Journal of the Fabry-Perot Observations.}
\label{JOFP_table}
\begin{tabular}{c|c|ccc|cc|cccc|cc}
\hline \hline        
Galaxy &  Date & \multicolumn{3}{c}{Filter}  & \multicolumn{2}{c}{Effective integration} & \multicolumn{4}{c}{Fabry-Perot} & \multicolumn{2}{c}{Sampling} \\
Name  & & $\lambda_c^{(4)}$ & FWHM$^{(5)}$ & T$_{max}^{(6)}$ &  t$_{exp}^{(7)}$ & t$_{Channel}^{(8)}$ & p$^{(9)}$ & FSR$^{(10)}$ & \textbf{\textit{F}}$^{(11)}$ & \textbf{\textit{R}}$^{(12)}$ & nch$^{(13)}$ & stp$^{(14)}$ \\ 
 &  & (\AA) & (\AA) & (\%)  & (min.) &  (min) & & (\kms) & & & & (\AA) \\
\hline 
NGC 628$^{(1)}$  & 18/11/2003 &   6598   &  18.2  &  73  & 149  & 2.33 & 899 & 333.47 & 23.6  & 21216 & 64  & 0.11  \\
NGC 925$^{(1)}$  & 11/02/2002 &   6584   &  15    &  75  & 132  & 2.75 & 765 & 391.88 & 16  & 12240 & 48  & 0.18  \\
NGC 2403$^{(1)}$  & 17/11/2002 &   6569   &  10    &  50  & 120  & 3.00 & 765 & 391.88 & 14 & 10710 & 40  & 0.21  \\
NGC 2798$^{(2)}$  & 04/04/2003 &   6608   &  16.2  &  69  & 96   & 2.00 & 899 & 333.47 & 16  & 14384 & 48  & 0.15  \\
NGC 2915$^{(3)}$  & 04/21/2004 &   6581   &  19.8  &  60  & 149  & 4.67 & 609 & 492.27 & 14.3  & 8580 & 32  & 0.34  \\
NGC 2976$^{(1)}$  & 30/01/2003 &   6581   &  19.8  &  60  & 184  & 3.83 & 899 & 333.47 & 16.5  & 14834 & 48  & 0.15  \\
NGC 3049$^{(2)}$  & 08/04/2003 &   6598   &  18.2  &  73  & 144  & 3.00 & 899 & 333.47 & 17.4  & 15643 & 48  & 0.15  \\
NGC 3031$^{(1)}$  & 06/02/2003 &   6581   &  19.8  &  60  & 272  & 5.55 & 899 & 333.47 & 14.3  & 12856 & 48  & 0.15  \\
UGC 5423$^{(2)}$ & 06/04/2003 &   6565   &  15    &  40  & 120  & 2.50 & 899 & 333.47 & 18.5  & 16632 & 48  & 0.15  \\
NGC 3184$^{(1)}$  & 18/02/2004 &   6584   &  15.5  &  74  & 162  & 3.37 & 765 & 391.88 & 17.6  & 13464 & 48  & 0.18  \\
NGC 3198$^{(1)}$  & 06/03/2003 &   6584   &  15.5  &  74  & 260  & 5.00 & 899 & 333.47 & 23  & 20976 & 52 & 0.14  \\
NGC 3521$^{(1)}$  & 19/02/2004 &   6584   &  15.5  &  74  & 120  & 2.50 & 765 & 391.88 & 16.0  & 12240 & 48  & 0.18  \\
NGC 3621$^{(3)}$  & 04/20/2004 &   6584   &  15.5  &  74  & 128  & 4.00 & 609 & 492.27 & 14.4  & 8640 & 32  & 0.34  \\
NGC 3938$^{(1)}$  & 11/03/2004 &   6584   &  15.5  &  74  & 128  & 2.66 & 765 & 391.88 & 16.8  & 12852 & 48  & 0.18  \\
NGC 4236$^{(1)}$  & 27/02/2004 &   6581   &  19.8  &  60  & 182  & 3.50 & 899 & 333.47 & 23  & 20977 & 52 & 0.14 \\
NGC 4321$^{(1)}$  & 25/02/2003 &   6608   &  16.2  &  69  & 260  & 5.00 & 899 & 333.47 & 23  & 20977 & 52 & 0.14  \\
NGC 4536$^{(1)}$  & 14/03/2004 &   6598   &  18.2  &  73  & 163  & 3.40 & 765 & 391.88 & 21.3& 16295 & 48 & 0.18  \\
NGC 4569$^{(1)}$  & 11/03/2002 &   6569   &  15.0  &  60  & 152  & 3.80 & 765 & 391.88 & 20.5& 15682 & 40 & 0.21 \\
NGC 4579$^{(1)}$  & 04/04/2002 &   6598   &  10.0  &  60  & 92   & 2.30 & 609 & 492.27 & 19.6& 14494 & 40 & 0.27 \\
NGC 4625$^{(2)}$  & 06/04/2003 &   6581   &  19.8  &  60  & 120  & 2.50 & 899 & 333.47 & 15.9  & 14294 & 48  & 0.15  \\
NGC 4725$^{(1)}$  & 19/02/2004 &   6584   &  15.5  &  74  & 120  & 2.50 & 765 & 391.88 & 18.7  & 14305 & 48  & 0.18  \\
NGC 5055$^{(1)}$  & 14/03/2004 &   6584   &  15.5  &  74  & 128  & 2.66 & 765 & 391.88 & 16.8  & 12852 & 48  & 0.18  \\
NGC 5194$^{(1)}$  & 18/05/2003 &   6581   &  19.8  &  60  & 246  & 5.14 & 899 & 333.47 & 20.7  & 18609 & 48  & 0.15  \\
NGC 5398$^{(3)}$  & 04/10/2004 &   6598   &  18.2  &  73  & 149  & 4.66 & 609 & 492.27 & 12.0  & 7308 & 32  & 0.34  \\
NGC 5713$^{(3)}$  & 04/13/2004 &   6608   &  16.2  &  69  & 150  & 6.25 & 609 & 492.27 & 9.5  & 5785 & 24  & 0.45  \\
IC 4710$^{(3)}$   & 04/15/2004 &   6598   &  18.2  &  73  & 48   & 2.00 & 609 & 492.27 & 8.8  & 5359 & 24  & 0.45  \\
NGC 6946$^{(1)}$  & 19/11/2002 &   6569   &  10    &  50   & 120  & 2.00 & 765 & 391.88 & 14    & 10710 & 40 & 0.21  \\
NGC 7331$^{(1)}$  & 03/11/2002 &   6584   &  15.5  &  74  & 174  & 3.62 & 765 & 391.88 & 15.9  & 12164 & 48  & 0.18  \\
\hline
\cr
\end{tabular}\\
\begin{flushleft}
$^{(1)}$ OMM : Observatoire du mont M\'egantic, Qu\'ebec, Canada. 1.6m telescope. \\
$^{(2)}$ CFHT : Canada-France-Hawaii Telescope, Hawaii, USA. 3.6m telescope.\\
$^{(3)}$ ESO : European Southern Observatory, La Silla, Chile, 3.6m telescope.\\
$^{(4)}$ $\lambda_c$ : non tilted filter central wavelength at 20$^{\circ}\mathrm{C}$.\\
$^{(5)}$ FWHM : non tilted filter Full Width Half Maximum at 20$^{\circ}\mathrm{C}$.\\
$^{(6)}$ T$_{max}$ : non tilted filter maximum transmission at $\lambda_c$ and at 20$^{\circ}\mathrm{C}$.\\
$^{(7)}$ t$_{exp}$ : total effective exposure time in minutes (Total exposure time * mean counting efficiency).\\
$^{(8)}$ t$_{channel}$ : total effective exposure time per channel in minutes (total exposure time per channel * mean counting efficiency).\\
$^{(9)}$ p : FP interference order at \ha.\\
$^{(10)}$ FSR : FP Free Spectral Range at \ha\, in \kms.\\
$^{(11)}$ \textbf{\textit{F}} : mean Finesse through the field of view.\\
$^{(12)}$ \textbf{\textit{R}} : spectral resolution ($\Delta\lambda/\lambda$) according to the computed \textit{Finesse}.\\
$^{(13)}$ nch : number of channels done by cycle in multiplex observations.\\
$^{(14)}$ stp : wavelength step in \AA.\\
\end{flushleft}
\end{table*}
\end{center}

\section{Observations}
\subsection{The sample}
\label{sample_section}

The SINGS sample as defined by \citeauthor{2003PASP..115..928K} (\citeyear{2003PASP..115..928K}) is composed of 75 nearby ($\Delta$ $<$ 30 Mpc, median of 9.5 Mpc, 
for $H_{0}=$ 70 \kms Mpc$^{-1}$) galaxies, covering a wide range in a 3D parameter space of 
physical properties:
\begin{itemize}
\item morphological type (E to Im), which is also correlated with the Star Formation Rate (SFR) per 
unit mass, gas fraction and bulge/disk ratio;
\item luminosity (IR-quiescent to luminous IR galaxies), which is also correlated with galaxy 
mass, internal velocity and mean metallicity;
\item FIR/optical ratio covering over 3 orders of magnitude, which is also correlated with
dust optical depth, dust temperature and inclination.
\end{itemize}
Roughly twelve galaxies were chosen in each RC3 type (E--S0, Sa--Sab, Sb--Sbc, Sc--Scd, Sd--Sm and 
Im--I0) which allows the coverage of a full combination of luminosity and infrared/optical ratio 
($5\times10^{5}L_{\odot} < L_{V} < 2\times10^{11}L_{\odot}$, $10^{7} L_{\odot} < L(IR) 
< 10^{11} L_{\odot}$ and $0.02 < L(IR)/L_{R} < 42$). Care was also taken to choose 
galaxies covering a wide range of other properties, such as nuclear activity, inclination, 
surface brightness, CO/\hI\, ratio, bar structure, spiral arm structure, isolated/interacting, 
group members, cluster members. Galaxies lying far from the Galactic plane were preferred to 
avoid a high density of foreground stars and galactic extinction.

From the 75 galaxies of the sample, only those which present \hII\, regions (star formation 
regions) can be observed in \ha\, in order to map their kinematics. Mainly, most early type galaxies 
(E to S0--Sa) lack \ha\, emission and could not be observed. Starting from Sb galaxies, it is 
usually possible to extract the \ha\, kinematics. Figure \ref{histo_figure} shows the morphological type distribution 
of the presented galaxies and highlights the observational bias caused by the lack of \ha\, emission 
in early-type galaxies. Also, Im--I0 galaxies are usually very small and could not be observed 
on the 1.6-m telescope at the Observatoire du mont M\'egantic. These galaxies need a 4 meters class telescope to be observed. The velocity maps and the observed galaxies' positions are shown in Figure \ref{great_map_figure}. This Figure also shows the relative sizes of the galaxies (in kpc). The basic galaxy parameters are presented in Table \ref{sample_table}. Table \ref{JOFP_table} gives the observation conditions and information for each galaxy.

\begin{figure*}
\includegraphics[width=\textwidth]{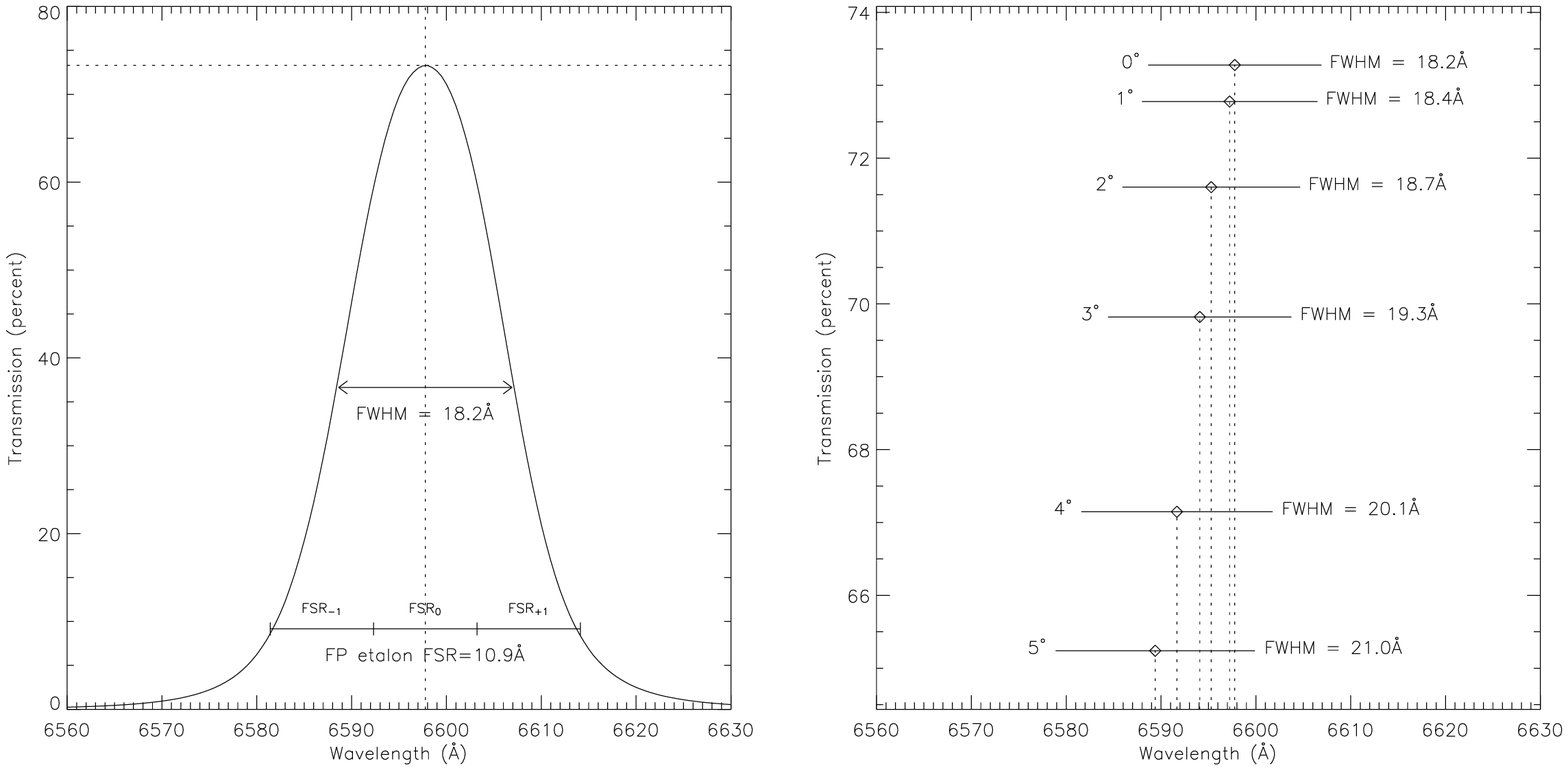}
\caption{Left panel shows a typical interference filter transmission curve (experimental data) 
at 20$^{\circ}\mathrm{C}$. As shown in the right panel, filters may be tilted by a few degrees to blue 
shift the peak transmission wavelength by a couple of \AA\, to allow it to be centered on the \ha\, emission of a 
galaxy, at the price of a broadened transmission curve and a decreased peak transmission. 
In the bottom of the left panel is shown how FP interferometer and interference filters 
must be chosen to avoid having too many Free Spectral Ranges to pass through the filter.}
\label{filtre_figure}
\end{figure*}

\subsection{The hardware}
\label{hardware_section}

All the observations were made with the integral field spectrometer, \FM. It consists of a narrow-band (typically 15\AA) interference filter, a FP interferometer and a photon counting camera, used as the imaging device. \FM\, is coupled to the respective focal reducer of the telescopes onto which it is attached. The focal reducer PANORAMIX is used at the 1.6-m telescope of the Observatoire du mont M\'egantic (OMM), CIGALE at the ESO (La Silla) 3.6-m telescope and MOS/FP at the CFH 3.6-m telescope. The effective focal ratio, pixel size and field of view are summarised in table \ref{fmcarac}. Interestingly, the pixel size obtained at the OMM (1.6\Sec), where most of the galaxies were observed, is a close match to the one achieved by the SPITZER's Infrared Array Camera (better than 2\Sec\, in the 3.6--8\um range). This is, however, pure coincidence.

The interference filter is used to ``select" the radial velocity range that will be 
observed. It is chosen to allow only the galaxy's \ha\, emission to pass through. 
Since most galaxies radial velocities span a maximum of $\pm$250\,$\mathrm{sin} i$\,\kms, a maximum Doppler 
shift of $\pm$5.5\AA\, is expected around the galaxy's redshifted emission. 
The interference filter used must allow this emission to pass through whilst being as 
narrow as possible to avoid too much sky background emission from reaching the detector. 
A collection of 23 filters having a FWHM of $\sim$ 15\AA, covering the red spectrum 
from 6565\AA\, to 6785\AA\, in steps of 10\AA\, was used to observe the galaxies in the 
sample. This set of filters allows galaxies with $-300$\kms $\lesssim v_{sys} \lesssim$ 10000\kms 
to be observed. Figure \ref{filtre_figure} shows a typical filter used for the observations. 
Centered at $\sim$6598\AA, it has a FWHM of 18\AA. By tilting the filter by a couple of degrees, it is possible to 
blue shift its central wavelength by a few \AA\, to allow the filter to be exactly centered on the 
galaxy's rest \ha\, emission. The filters must also be chosen according to the expected outdoor 
temperature as they are typically blue shifted as temperature goes down. This shift is of the 
order of $-0.2$\AA$\cdot\mathrm{K}^{-1}$. As filters age, they also tend to blue shift. They must be scanned regularly to keep a good knowledge of their characteristics.

The FP interferometer is chosen to allow most of the \ha\, emission of the galaxy to be visible in 
a single Free Spectral Range (FSR). The interference orders of the FP interferometers used vary 
from p=609 (FSR = 10.93\AA) to p=899 (FSR = 7.3\AA) at rest \ha. Another parameter that has to be 
taken into account is the FP \textit{Finesse}. The Finesse, F, is a dimensionless value expressing 
the spectral resolution, R, of the etalon such as 
$$R = \frac{F\lambda_0}{FSR} = \frac{\Delta\lambda}{\lambda}$$
and
$$FSR = \frac{\lambda_0}{p}.$$
To properly sample the light that comes out of the FP, the number of channels scanned must be at least 2.2 times 
(Nyquist) the Finesse. Many different factors affect the Finesse: the reflectivity of the reflecting plates of the FP, the optical and mechanical properties of the surfaces (usually polished to $< \frac{\lambda}{100}$), the temperature, the humidity, and the observational setup (the parallelism, the accuracy of the focus of the focal reducer). In fact, the environmental effects and the observational setup should not affect the Finesse, but, as shown by Table \ref{tableFP}, the scatter in Finesses obtained must be explained by these phenomena. Finesses achieved for the observations 
presented in this paper range from 8 to 20. A total of 3 different FP etalons were used for the observations.
The interference filter and PF interferometer must be chosen so that no more that 3 FSRs of the etalon pass through the filter, as shown in 
figure \ref{filtre_figure}.

The photon counting camera used for this survey is based on a GaAs Hamamatsu photocathode coupled with a 
Dalsa commercial CCD. The photocathode has a quantum efficiency of $\sim$28\% at \ha\, and the 
CCD has 1024$\times$1024 12.5\um\, square pixels. The CCD was operated in its low spatial 
resolution mode, where pixels are binned 2$\times$2, and at a frame rate of 40 frames per seconds. The effective pixel sizes on the sky are given 
in Table \ref{fmcarac}. The photon counting camera is an essential tool to achieve such 
a survey. Its ability to rapidly scan the FP interferometer allows the photometrical conditions' 
variations to be averaged out. For comparison, in CCD observations, each FP channel must be 
observed for at least 5 contiguous minutes to make sure that the read-out noise of the CCD does not mask
the weak galaxy's signal. This means that photometrical conditions must not significantly 
change for $\sim$4 hours with CCD observations (given that 48 FP channels are scanned). In photon counting, channels are observed 
for 5 to 15 seconds and cycles are looped every 4 to 15 minutes. Many cycles are made 
during an observation. Signal-to-noise ratio (SNR) estimations can be made throughout the observations and the 
observer can decide when to stop the integration.
Since the calibration lamp is pretty strong as compared to galaxies' fluxes, the calibrations
are done in analog mode (non photon counting). Typically, calibration channels are 
integrated 1 second each.

\subsection{Observing runs}
\label{obs_runs_section}

The observations of the sample presented here were spread over ten observing runs taking 
place in the last two years. Eight runs took place at the Observatoire du mont Megantic 1.6-m
telescope (OMM), where \FM\, is a permanent instrument. \FM\, was taken to the 
Canada-France-Hawaii 3.6-m Telescope (CFHT) and the European Southern Observatory 3.6-m 
telescope (ESO/La Silla) as a visitor 
instrument for the other two runs of this survey. The large field of view available 
at the OMM permitted the observation of large galaxies (NGC 5194, NGC 3031...) in one single 
field. Smaller galaxies visible in the northern hemisphere were observed at the CFHT. A 
single observing run took place in the southern hemisphere on the ESO/La Silla telescope 
in April 2004. Few galaxies of the SINGS survey were visible at that time, resulting in a poor coverage of 
galaxies at negative declinations.


\section{Data reduction}

Raw observational data consist of many data files that contain photons' positions for every 
cycle/channel duo. If cycling is done with, say, 15 seconds per channel integration time, 
one file is created every 15 seconds. Data reduction consists of the following steps:
\begin{itemize}
\item integration of the raw data files into an interferogram data cube (3D data cube sliced for every FP channel);
\item phase correction of the interferograms to create wavelength-sorted data cubes (3D data cube
sliced for every wavelength interval);
\item hanning spectral smoothing;
\item sky emission removal;
\item spatial binning/smoothing;
\item extraction of radial velocity maps from emission line positions;
\item addition of the astrometric information (World Coordinate System information) to the data files;
\item kinematical information extraction.
\end{itemize}

All the reduction was performed with IDL routines inspired by the ADHOCw reduction package 
(\url{http://www-obs.cnrs-mrs.fr/adhoc/adhoc.html}), except for the last two steps, in which cases third party software was
used (see respective sections for more details). The IDL reduction package was written 
to allow more flexibility of the data reduction, 
such as telescope guiding error correction, better removal of the sky background emission through sky cube fitting, stronger 
emission line barycenter determination algorithm, etc. The newly introduced reduction routines 
are summarized here. More information on this reduction package will be available in a forthcoming paper (\citeauthor{2005nowhere.2..2A} \citeyear{2005nowhere.2..2A}).

\subsection{Raw data integration and wavelength map creation}

Files stored during an observation are made of the position of every photon that fell on the 
detector for a given FP position. The observed wavelength through a FP 
etalon is given by
$$p\lambda=2ne\cos\theta\,,$$
where $p$ is the interference order at $\lambda_0$ (\ha\, or 6562.78\AA\, for all observations 
presented in this paper), $n$ the index of the medium, $e$ the distance between the plates of the FP and $\theta$ the incidence angle 
on the FP. For a given gap $e$ and index $n$, every $\theta$ in a single channel is exposed 
to a different wavelength, which is to say that different pixels see different wavelengths. Thus, a phase calibration
must be applied to transform raw interferograms into wavelength-sorted data cubes. This calibration is obtained by 
scanning a narrow Ne emission line. The bright line at 6598.95\AA\, has been chosen since 
it falls close to the red-shifted \ha\, emission lines observed, limiting the phase shift
in the dielectric layer of the FP. This calibration must be 
done in exactly the same conditions as the observations. It is thus convenient to acquire 
this calibration data just before starting an integration, when the telescope is already 
on the object. Since these calibrations are done with the camera in analog mode 
(as compared to photon counting mode), the overhead is very small 
(of the order of a minute). This allows one to perform a calibration at the beginning 
and at the end of an integration and to sum them in order to average tiny variations of the 
FP etalon during the integration.

\begin{table}
\caption{\FM\ characteristics on various telescopes.} 
\begin{center}
\begin{tabular}{cccccc}
\hline \hline
Telescope & D & $F/D$ & Pixel size & FOV & FOV$_{unvign}$ \\
Name  & m$^{(1)}$ & $^{(2)}$ & (\Arc)$^{(3)}$ & (\Min)$^{(4)}$ & (\Min)$^{(5)}$ \\
\hline
CFHT & 3.6 & 2.96 & 0.48 & 5.83 & 3.91 \\
ESO  & 3.6 & 3.44 & 0.42 & 5.01 & 5.01 \\
OMM  & 1.6 & 2.00 & 1.61 & 19.43 & 19.43 \\
\hline
\cr
\end{tabular}
\end{center}
\begin{flushleft}
$^{(1)}$D : telescope diameter in meter.\\
$^{(2)}F/D$ : focal length over telescope diameter ratio calculated from the 
effective pixel size on the sky.\\ 
$^{(3)}$Pixel size : pixel size after binning 2$\times$2, the original GaAs system providing 1024$\times$1024 pixels 
of 12.5\um.\\ 
$^{(4)}$FOV : diagonal Field Of View of the detector.\\ 
$^{(5)}$FOV$_{unvign}$ : unvignetted (usable) field of view.
\end{flushleft}
\label{fmcarac}
\end{table}

From this calibration data, a phase map is created. The phase map provides the shift that 
has to be applied in the spectral dimension to every pixel of the interferogram cube 
to bring all channels to the same wavelength. By applying this phase map to the interferogram data cube, the 
wavelength-sorted data cube is created. An uncertainty still remains on the zero 
point of the velocity scale since the observed 
wavelength is only known modulo the FSR of the interferometer
$$FSR\equiv\Delta\lambda=\frac{\lambda_0}{p}.$$
This uncertainty is removed by means of comparisons with long-slit spectroscopy or 21-cm 
\hI\, data.

The computed wavelength of every slice of the resulting (calibrated) data cube is not absolute. 
When the observed wavelength is far from the calibration wavelength, the difference will 
increase. This is caused mainly by the semi reflective, high Finesse coating of the FP 
etalon, which is hard to model. Absolute calibrations (in development), done at the same 
scanning wavelength would be a way of getting rid of these differences. Absolute post-calibration 
using observed sky spectrum emission lines in the data cube, which must include compensation 
for the interference filter transmission curve, has been worked on but requires more testing to prove 
its accuracy. Nevertheless, \textit{relative} wavelength measurements are accurate to a 
fraction of a channel ($<$ 0.05\AA ) over the field. This leads to line of sight 
velocity (LOSV) measurement errors of less than 3\kms .

Since the observational data are split in files representing a maximum of 15 seconds 
integration, it is possible to correct slight telescope guiding errors that could
occur throughout the entire integration (typically 2--3 hours). This correction 
must be made at the same time as the raw data integration. But, as a spatial translation 
in the interferogram domain will induce a wavelength shift in the spectral domain, 
the phase calibration must be applied at the same time as the raw integration. 
Results presented in this paper have undergone the phase calibration at the 
same time as the raw data integration in order to render guiding error corrections possible.

\subsection{Sky emission removal}

The sky emission in the neighborhood of \ha\ at rest, caused by geocoronal OH molecules is often stronger 
than the galaxy's diffuse \ha\, emission. It is thus highly important to properly remove this emission 
prior to the extraction of the radial velocity map. As opposed to the ADHOCw package, 
where the OH emission lines are removed by the subtraction of a single integrated profile taken from 
 user-selected sky regions, the data presented in this paper have been processed differently. Due to inhomogeneities in the spatial and spectral responses of the interference filter, the subtraction of a single spectrum can lead to positive and 
negative residuals in the data cube, where not enough or too much of the spectrum were 
subtracted. To avoid having to deal with such issues, and to improve the galaxies' 
signal coverage, a sky cube has been reconstructed by taking single pixel's spectra in the sky 
dominated regions and interpolating (or extrapolating) it in the galaxy's dominated spectra. 
This sky cube was then subtracted from the data cube. This proved to be a much better procedure. For best results, the surface covered by the sky dominated regions and the galaxy dominated regions should be in a $\sim$1:1 ratio. 

\subsection{Adaptive spatial binning and smoothing}

\begin{table}
\caption{FP interferometers characteristics} 
\begin{center}
\begin{tabular}{ccccc}
\hline \hline
Interferometer & p & FSR & F & R\\ 
\hline
1 & 899 & 333.47 & 15.6 -- 23.6 & 14294 -- 21216 \\
2 & 765 & 391.88 & 14.0 -- 21.3 & 10710 -- 16295 \\
3 & 609 & 492.27 & 8.8 -- 19.6  & 5259 -- 14494 \\
\hline
\cr
\end{tabular}
\end{center}
\label{tableFP}
\end{table}

In order to allow the LOSV of a galaxy's region to be properly determined, a minimum SNR is required. \citeauthor{2003MNRAS.342..345C} (\citeyear{2003MNRAS.342..345C})
used 2D Voronoi tessellations to create uniform SNR velocity maps of Sauron data. Starting 
from this work, an adaptive binning algorithm was developed for FP 3D data cubes. 
The main difference between the two algorithms is the way by which pixels are accreted into 
bins. Where Cappellari \& Copin compute the SNR of a bin by means of
$$SNR = \frac{\sum_{i}Signal_{i}}{\sqrt{\sum_{i}Noise_{i}}}\,,$$
where $Signal_{i}$ and $Noise_{i}$ are precomputed signal and noise value of the spectra that will be binned, 
the SNR of the spectra presented in this paper have been recomputed each time a new spectrum was added 
to a bin. In short, each time a spectrum was added, it was summed with the other spectra 
of the bin and a new SNR was recomputed by means of
$$SNR = \frac{N}{\sqrt{N+\sigma^{2}}}\,,$$
where $\sigma$ is the dispersion of the continuum of the spectrum and $N$ the number of photons 
composing the emission line located above the continuum. Typically, a target SNR of 5 has been 
used. After the binning process, a Delaunay triangulation algorithm was used to smooth bins 
of each channel of the data cubes.

This smoothing method has been preferred to the fixed-sized kernel convolution (such as a
6$\times$6 gaussian). Adaptive spatial binning allows the spatial resolution to be kept in high SNR 
regions while still providing large signal coverage in low SNR regions.

\subsection{Radial velocity map extraction}
\label{rvmap_section}

Radial velocity, monochromatic, continuum and dispersion maps are extracted with a single 
emission line detection algorithm. This algorithm is based on barycenter computation. The central position of the emission line 
is computed with photons above the continuum that constitute the emission line. The radial 
velocity is then computed as
$$v_{obs} = \left(\frac{\left(\frac{\lambda_{obs}}{\lambda_{0}}\right)^{2}-1}{\left(\frac{\lambda_{obs}}{\lambda_{0}}\right)^{2}+1} + 1\right) * c + corr\,,$$
where $\lambda_{obs}$ is the emission line computed barycenter wavelength, $\lambda_{0}$ the 
rest wavelength (here, \ha), $c$ the speed of light in vacuum and $corr$ the heliocentric 
velocity correction computed for the time of the observation.

In the case where more than one velocity component are present in the spectrum, only the strongest emission line will be taken into account. When two emission lines are spectrally close and have comparable amplitudes, they may be taken as a single one having a larger velocity dispersion. More details are available in \citeauthor{2005nowhere.2..2A} (\citeyear{2005nowhere.2..2A}).

\subsection{Astrometry}

Astrometric information was attached to the processed files by using the task \verb|koords| in 
the \textsc{Karma} package (Gooch, 1996). Right ascension, declination, pixel size and 
field rotation information are embedded in all files and permit their easy comparison 
with other survey images (DSS, 2MASS, SPITZER). To do so, original ADHOCw files 
types (ad2, ad3) were modified to allow these data to be stored. This is also an important 
step since the position angle (\PA) of the major axis, whose determination is explained in the next section, 
is field orientation dependent. Moreover, the astrometric information is necessary to combine the kinematical data to the SINGS and ancillary multi wavelength surveys.

\begin{table}
\caption{Photometrical and kinematical parameters}
\begin{center}
  \begin{tabular}{cccccc}
  \hline\hline
   Galaxy  &   \multicolumn{2}{c}{Photometrical$^{(1)}$} & \multicolumn{2}{c}{Kinematical}  \\
   name &   \PA(\Deg) & Incl.(\Deg) & \PA(\Deg)& Incl.(\Deg)    \\
 \hline
NGC 628   &  25       &  24   &   26.4$\pm$2.4           &   21.5$\pm$4.5 \\
NGC 925   & 102       & 56    & 105.0$\pm$1.0            &  50.0$\pm$1.5 \\
NGC 2403  & 127       & 56    & 125.0$\pm$1.0            &  60.0$\pm$2.0\\
NGC 2798  &  160      &  68   &   $-^{(2)}$              &   $-^{(2)}$ \\
NGC 2915  &  129      &  59   &   $-^{(2)}$              &   $-^{(2)}$ \\
NGC 2976  &  323$^{(3)}$      &  63   &   323.5$\pm$3.5  &   70.2$\pm$4 \\
NGC 3049  &  25       &  49   &   $-^{(2)}$              & $-^{(2)}$  \\ 
NGC 3031  &  337$^{(3)}$      &   58  &   332.9$\pm$1.2  &   62.4$\pm$1 \\
UGC 5423  &  320$^{(3)}$      &  49   &   320.5$\pm$5    &   58.8$\pm$6 \\
NGC 3184  &  135      &  21   &   176.4$\pm$4$^{(4)}$    &   16.7$\pm$1.1 \\
NGC 3198  &  35       &  67   &    33.9$\pm$0.3          &  69.8$\pm$0.8\\ 
NGC 3521  &  343$^{(3)}$      &  62   &   342.0$\pm$1.1  &   66.7$\pm$2 \\
NGC 3621  &  339$^{(3)}$      &  55   &   342.5$\pm$2.9  &   65.2$\pm$4 \\
NGC 3938  &  151$^{(3)(5)}$&  24   &   199.5$\pm$5$^{(4)}$    &   7.7$\pm$3.2 \\
NGC 4236  &  162      & 71    &   156.1$\pm$1.6          & 76.1$\pm$0.7\\
NGC 4321  &  30       & 32    &    27.0$\pm$1.0          &  31.7$\pm$0.7\\
NGC 4536  &  310$^{(3)}$      &   65  &   300$\pm$2      &   68$\pm$3 \\
NGC 4569  &  23       &  63   &   24.7$\pm$4.4           &  66.7$\pm$5 \\
NGC 4579  &  95       &  37   &   89.5$\pm$4.5           &   45.7$\pm$7.3 \\
NGC 4625  &  162$^{(3)(5)}$&  29   &   126$\pm$5$^{(4)}$     &   35.9$\pm$6 \\
NGC 4725  &  35       &  45   &   30.3$\pm$1.6           &   50.8$\pm$2.1 \\
NGC 5055  &  105      &  55   &   98.0$\pm$1.9           &   63$\pm$2 \\
NGC 5194  &  163      &  52   &   169.0$\pm$4.2          &   47$\pm$5 \\
NGC 5398  &  172      &  53   &   $-^{(2)}$              &   $-^{(2)}$ \\
NGC 5713  &  190$^{(3)}$      &  27   &   203$\pm$5.8    &   33$\pm$4.8 \\
IC 4710   &  5        &  39   &   $-^{(2)}$              &   $-^{(2)}$ \\
NGC 6946  & 250$^{(3)(5)}$ &  32   & 239.0$\pm$1.0       & 38.4$\pm$3.0\\
NGC 7331  & 171       &  69   &   165$\pm$1.2            &   78.1$\pm$2.7 \\
\hline \\
\end{tabular}
\end{center}
\begin{flushleft}
$^{(1)}$Photometrical parameters taken from the RC3 catalog.\\
$^{(2)}$It is useless or impossible to determine the kinematical parameters for this galaxy due either to the lack of large-scale rotation or the interaction with another galaxy.\\
$^{(3)}$A 180\Deg\, rotation was applied to the photometrical PA in order to be able to compare it with the kinematical PA.\\ 
$^{(4)}$The photometrical and kinematical \PA\, do not agree even with a 180\Deg\, rotation. This usually happens for face-on galaxies.\\
$^{(5)}$This photometrical parameter was not available in the RC3 database and was taken from the 2MASS Large Galaxies Atlas.\\
\label{kinematics}
\end{flushleft}
\end{table}

\subsection{Kinematical parameters}
\label{kinematics_section}
The \textsc{Rotcur} routine in the \textsc{Gipsy} package was used to find the kinematical 
parameters of the galaxies studied. \textsc{Rotcur} works by fitting tilted ring models to 
velocity information found in the velocity maps. $V_{obs}$ is obtained using the following equation:
$$V_{obs} = V_{sys} + V_{rot}\left(R\right)\cos\theta\sin i + V_{exp}\left(R\right)\sin\theta\sin i\,,$$
where $V_{sys}$ is the systemic velocity of the system studied, $V_{rot}$ is the rotation 
velocity, $V_{exp}$ the expansion (non-circular) velocity, $i$ the inclination of the ring 
and $R$ and $\theta$ the polar coordinates in the plane of the galaxy. The same procedure 
was used for all the galaxies of the sample. To allow for a good sampling frequency, the 
ring width used was always at least 3 times the pixel size.

The extraction of the kinematical parameters is done as follows. The photometrical parameters 
of the galaxies (position angle ($\mathrm{PA}$), inclination ($i$)) are taken from the RC3 catalog as initial parameters. The photometrical inclination is calculated as $I = \cos^{-1}\left(10^{-R_{25}}\right)$.
The rough rotational center of the galaxies is taken by looking at the continuum maps extracted 
at section \ref{rvmap_section} and taking the point where the continuum is the highest 
near the center of the galaxy (usually obvious for spiral galaxies, trickier for irregular, low surface brightness and
distorted ones). The starting point of the systemic velocity is the one used to select 
the interference filters, as explained in section \ref{hardware_section}.

The initial photometrical PA must be sometimes corrected ($\pm$180\Deg) to properly represent 
the kinematical $\mathrm{PA}$, that is, the angle from North Eastwards to the receding side of the 
galaxy. Also, \textsc{Gipsy} does not take into account the field rotation of the supplied 
radial velocity map. The starting \PA\, is adjusted in order to reflect this.

First, \textsc{Rotcur} is run to find the real kinematical center of the galaxy and its 
systemic velocity by fixing both $\mathrm{PA}$ and $i$ and leaving the other parameters 
free. The analysis of the output of \textsc{Rotcur} is made with IDL routines that permit 
the extraction of statistics on the free parameters, such as median, mean, error weighted 
mean, standard deviation and linear fit. If the output is too noisy, \textsc{Rotcur} was run 
again, starting with another set of initial parameters, until the median of the absolute residual of the computed model was below 10\kms. Then, having properly determined the kinematical center and the systemic 
velocity, the real kinematical $\mathrm{PA}$ and inclination were set as free parameters, fixing 
all the others. Finally, having found the five kinematical parameters (x$_c$, y$_c$, $V_{sys}$, 
$\mathrm{PA}$ and $i$), they were fixed and \textsc{Rotcur} was run again to find $V_{rot}$. It was 
decided to use fixed values of $\mathrm{PA}$ and $i$ across the whole galaxy as disks are rarely 
warped inside their optical part. Warps are mainly seen in disks for $R > R_{opt}$. However, the PA is observed to vary as function of radius for some galaxies (e.g. \mbox{NGC 4579}). The PA value is thus chosen in part of the disk where it reaches an almost constant value. The errors on the kinematical parameters found can thus be artificially lowered since the non-axisymmetric parts of the galaxies are discarded from the fit. For barred galaxies, this involved restricting the fit outside the bar. For non-barred galaxies, the outer parts were removed from the fit when the errors were larger than four times the means error.


\section{Results}
The kinematical parameters found by means of the method described in section 
\ref{kinematics_section} are presented in table \ref{kinematics}.

In Appendix \ref{images_results}, for each galaxy of the sample, the XDSS blue image, the 
SPITZER 3.6\um image, the \ha\, monochromatic image, where the continuum has been suppressed, and the RV map are provided. 
A Position-Velocity (PV) diagram is given when it was possible to extract the
kinematical parameters from the radial velocity map. The red line superposed to the PV diagram is the velocity of the major axis of the galaxy's model reconstructed from its rotation curve. All images for a given galaxy have the same angular scale. Blue, IR and \ha\, monochromatic images have a logarithmic intensity scale. The color scales of the radial velocity maps and the PV diagrams are linear.

Since the calibrations and the observations are most of the time done through different interference filters, the presented monochromatic maps could not be flux calibrated.

The rotation curves, extracted as explained in Section \ref{kinematics_section}, are presented in Appendix \ref{rotation_curves}. The errors given for the velocity points are the difference between the approaching and the receding side of the galaxy or the formal error given by rotcur, whichever is the largest. We think that this is a better estimate than using directly the formal error of \verb|rotcur| since this will take into account possible asymmetries between the two sides of the galaxies.


The rotation curves are available in electronic form at \url{http://www.astro.umontreal.ca/fantomm/sings/rotation_curves.htm}.


\begin{figure*}
\begin{center}
  \includegraphics[width=\textwidth]{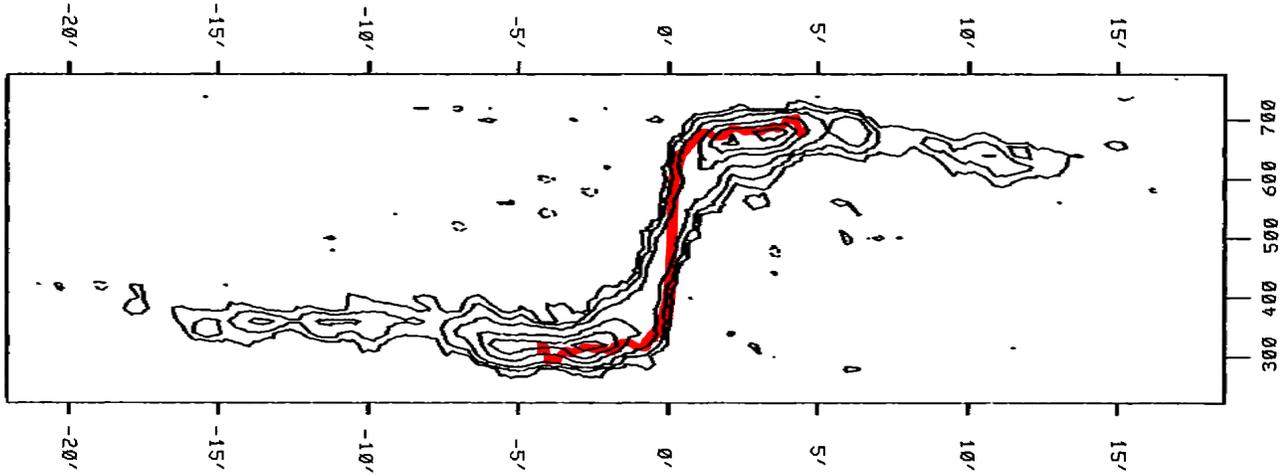}\\
\end{center}
\caption{\hI\, data of NGC 5055 taken from \citeauthor{1986A&AS...66..505W}, 
\citeyear{1986A&AS...66..505W}, and \ha\, rotation curve superposed on it (red line). 
This figure clearly shows the effect of beam smearing on \hI\, data and the need 
for \ha\, data to resolve the kinematics at the center of galaxies. In \hI, the 
beam Full Width Half Power (FWHP) was 49\Sec x 73\Sec while the \ha\, resolution 
is 1.6\Sec x 1.6\Sec}
\label{ngc5055_h1_ha}
\end{figure*}

\section{Discussion}
\label{section_discussion}

In this section, we discuss the advantages and limitations of \ha\, kinematical observations using integral field spectroscopy over other kinematical observational methods.

\subsection{Better resolution of the inner rotation curve}
Of course, the signal coverage of the \ha\, observations is less extended than for the \hI\, observations 
but it allows to resolve the rising part of the rotation curve with greater precision. 
Figure \ref{ngc5055_h1_ha} shows a 21-cm PV diagram of NGC 5055 taken from 
\citeauthor{1986A&AS...66..505W} (\citeyear{1986A&AS...66..505W}). From the 21-cm data 
alone, one can calculate that the maximum velocity gradient at the center of the galaxy is of the order of $\mathrm{\sim45\,kms^{-1} kpc^{-1}}$. The red line shows the rotation curve derived from the \ha\, data, which has a maximum gradient of $\mathrm{\sim300\,kms^{-1} kpc^{-1}}$. This shows how the 21-cm data are affected by beam smearing. For \hI, the beam 
Full Width Half Power (FWHP) is 49\Sec x 73\Sec\, while the \ha\, resolution 
is 1.6\Sec x 1.6\Sec.

The rising part of the rotation curve is crucial in the 
determination of the dark-to-luminous mass ratio (\citeauthor{1999AJ....118.2123B}, 
\citeyear{1999AJ....118.2123B}) and in the determination of the mass model parameters. 
For the sake of comparison, maximum achievable 21-cm beam width at the VLA in B 
configuration is $\sim$4\Sec\, usable only for the strongest emitting galaxies, 
while it is $\sim$12.5\Sec\, in C configuration, where sensitivity is high enough to 
observe the weak galaxies' signal. \mbox{NGC 5055} harbors double emission lines in its center, as noted 
by \citeauthor{2004A&A...420..147B} (\citeyear{2004A&A...420..147B}). 
Resolving such details needs both the high spatial and spectral resolution 
of the integral field spectroscopy used throughout this study.

\subsection{Better determination of the orientation parameters}

One of the main advantage of the determination of galaxies' kinematics using 
integral field spectroscopy at \ha\, over long slit spectroscopy is that there is no \textit{a priori} knowledge needed on the galaxy apart from its systemic 
velocity (which is usually well known within a $\pm$50\kms range for nearby galaxies, which is enough for accurate observations). In long slit spectroscopic observations, the \PA\, of the galaxy must be known as the slit must lie on the major axis. The $\mathrm{PA}$ is thus usually determined by fitting ellipses to the optical isophotes. As Table \ref{kinematics} shows, there is sometimes a great discrepancy between published photometrical and kinematical parameters. 

The effect can be well illustrated with the
nearly face-on ($\mathrm{i=17^{\circ}}$) galaxy NGC 3184. 
The photometrical PA of 135\Deg\, is 41\Deg\, off from the kinematical \PA\, found. 
This \PA\, error can lead to a substantial underestimate of the rotation velocities for 
highly inclined galaxies. Also, for some galaxies, the kinematical center is 
not superposed on the photometrical center. This can also lead to large errors 
on the rotational velocities.

\begin{figure*}
\begin{center}
  \includegraphics[width=\textwidth]{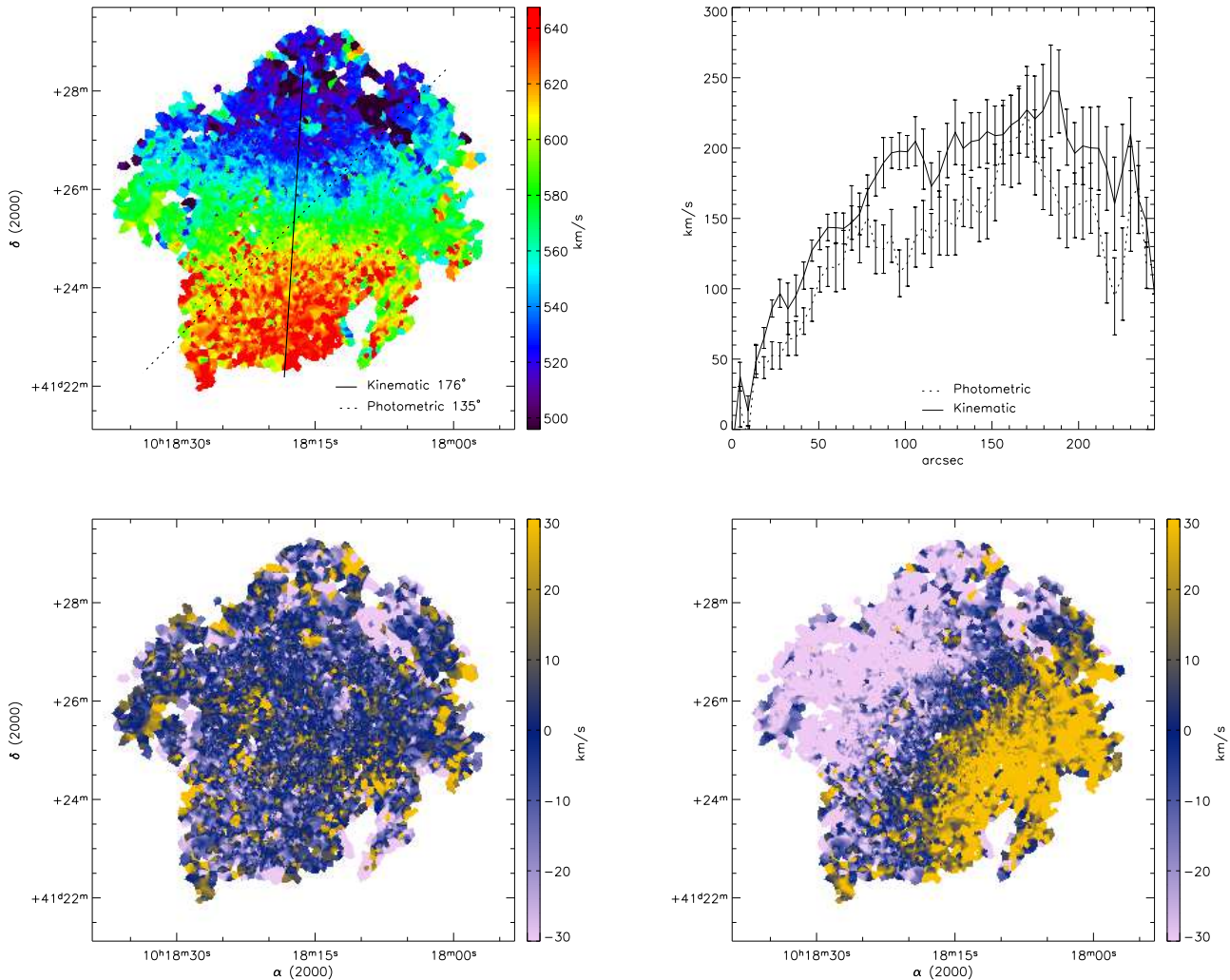}\\
\end{center}
\caption{Difference between kinematical and photometrical parameters for NGC 3184. 
\textbf{Top left} : \ha\, velocity field showing kinematical and photometrical PA. 
\textbf{Top right} : Rotation curves extracted from the velocity fields using kinematical 
and photometrical parameters as if long slit spectroscopy was used for the observations. 
The RMS error of the photometrical rotation curve is 38.4\kms. There is nearly a 30\% error 
around 100\Sec. \textbf{Bottom left} : Residual map of a model built using the kinematical 
parameters. \textbf{Bottom right} : Residual map of a model built using the photometrical 
parameters.}
\label{ngc3184_photo_kine}
\end{figure*}

Figure \ref{ngc3184_photo_kine} shows the effect of using the photometrical \PA\, to 
observe this galaxy in long slit spectroscopy. The resulting rotation curve is less 
steep in the center of the galaxy and the maximum rotational velocity is underestimated 
(RMS error is 38.4\kms), which affects the mass models used to determine the dark 
matter content. Using integral field spectroscopy, the kinematical \PA\, is 
determined \textit{a posteriori} and it does not affect the quality of the data 
gathered. The figure also presents the residuals of the velocity field models built from the 
rotation curves extracted. It clearly shows that would the kinematical data have 
been gathered from long slit observations using the photometrical PA, the results 
would have been totally erroneous. 

On the other hand, it should be considered that the difference between the photometrical and kinematical PA may be partially due to the fact that the photometrical \PA\, would have been more accurately determined by using deep IR images (such as SPITZER images), which shows the old stellar population and is less affected by structural patterns.

\subsection{Observation of non-axisymmetric motions}

Barred galaxies lead to a specific problem since errors on the determination of the PA of the disk can be induced by the bar. For instance, the galaxy \mbox{NGC 3049} is totally dominated by a bar and its associated non-axisymmetric motions. 
By taking a look at the blue and IR images of Figure \ref{figure_ngc3049}, it could be 
thought, at first, that the \PA\, is somewhere around 20\Deg. However, the isovelocity 
contours suggest that the \PA\, is more like 60\Deg. This problem is caused by the 
lack of kinematical information outside the bar dominated region of the galaxy. 
In this case, \hI\, data 
would be necessary to resolve the kinematics outside the bar, since the global kinematical 
parameters must be extracted from the axisymmetric portion of the galaxy.

This galaxy demonstrates clearly the advantage of 2D velocity fields over 1D long-slit data. 
2D velocity fields allow to disentangle circular from radial motions while they would be confused in 
long slit data. Integral field spectroscopy makes it possible to study more thoroughly 
non circular motions in galaxies, such as in \citeauthor{2005MNRAS.360.1201H} (\citeyear{2005MNRAS.360.1201H}), where an in-depth study of barred galaxies is done.

\subsection{Observation of highly inclined galaxies}
\begin{figure*}
\begin{center}
  \includegraphics[width=0.75\textwidth]{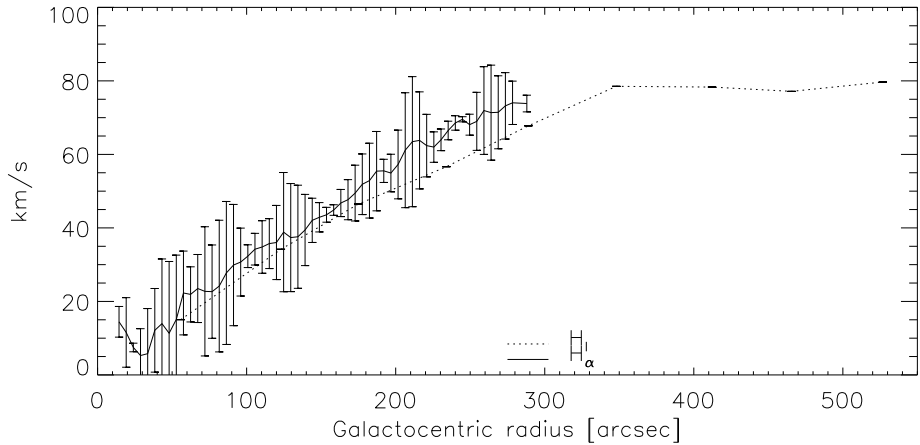}\\
\end{center}
\caption{NGC 4236 rotations curves obtained in \ha\, and \hI. The \hI\, data are taken from \citeauthor{1975A&A....45..259H} (\citeyear{1975A&A....45..259H}).}
\label{ngc4236_rc}
\end{figure*}

It has been said that \ha\, could not resolve the kinematics for highly inclined 
galaxies since the gas cannot be considered optically thin at this wavelength (\citeauthor{1981ApJ...250...79G} \citeyear{1981ApJ...250...79G}). 
But, the case of NGC 4236, whose inclination is 76\Deg\,, shows that as long as 
the galaxy is not perfectly edge-on, the major axis of the galaxy is visible and 
the kinematics of the disk can be resolved. Figure \ref{ngc4236_rc} shows how 
the \ha\, and \hI\, kinematics agree. As opposed to the case of \mbox{NGC 5055}, we can compare the \ha\, kinematics to the \hI\, kinematics since the inner part of the rotation curve is less shallow. However, since the signal coverage is less extended for \ha\, than it is for \hI, the 
flat part of the rotation curve is missing from the \ha\, data. Also, as stated by \citeauthor{1992ApJ...400L..21B} (\citeyear{1992ApJ...400L..21B}), as spiral galaxies can be considered optically thin at least for the outer part of the visible disk ($\mathrm{R > 0.5R_{25}}$), \ha\, observations can be used to resolve their kinematics.

\section{Conclusion}
\label{section_conclusion}
The \ha\, kinematics of \NN\, galaxies of the SINGS survey were presented in this paper. The observations were made with \FM, an integral field FP spectrometer and a photon counting camera. The raw data obtained at the telescope were processed through a new pipeline, an adaptive binning algorithm has been applied to achieve optimal signal-to-noise ratio and the radial velocity maps were finally extracted from the data cubes using a selective intensity weighted mean algorithm. Kinematical parameters were computed using a tilted ring model and most of them agreed within an acceptable error range with the photometrical parameters, except for a few problematic galaxies. It has been shown that high spatial resolution data is essential for mapping the velocity gradient at the centres of galaxies. The advantages of integral field spectroscopy over long slit spectroscopy were also presented.

The aim of this paper is to provide accurate optical kinematical data for the galaxies of the SINGS survey. These data will be used in a forthcoming paper to present rotation curves and mass models of the non-barred galaxies (Nicol et al. in preparation). The data of some of the barred galaxies of the SINGS sample that are also part of the \BB\, survey were used to derive the bar pattern speeds using the Tremaine--Weinberg method (\citeauthor{1984ApJ...282L...5T} \citeyear{1984ApJ...282L...5T}) in a paper presented by \citeauthor{2005ApJ...632..253H} (\citeyear{2005ApJ...632..253H}). These data might also be used to allow for some of the galactic star formation models to include kinematical data and thus try to determine what is the exact role of rotation in the star formation processes on a galactic scale. 

The \ha\, kinematical data for all the observable SINGS galaxies will be made available to the community as soon as all the galaxies have been observed.


\section*{Acknowledgements}
          
We thank Jean-Luc Gach, Philippe Ballard, Olivia Garrido, Jacques Boulesteix and
Olivier Boissin from the OAMP, for their help and support at the different stages of this work.  
Many thanks also to Bernard Malenfant and Ghislain Turcotte, from the OMM,
Pierre Martin and the CFHT staff as well as the ESO 3.6-m telescope team who helped us a lot in making the observing runs a success. We also want to thank the anonymous referee for his valuable comments. 
The \FM\, project has been carried out by the Laboratoire d'Astrophysique 
Exp\'erimentale (\textbf{\texttt{LAE}}) of the Universit\'e de Montr\'eal using a 
grant from the Canadian Foundation for Innovation and the Minist\`ere de l'Education 
du Qu\'ebec. 
This project made use of the LEDA database: \textsf{http://leda.univ-lyon1.fr/}.

\appendix

\section[]{Description of the individual galaxies}

A brief description of the structures observed in the \ha\, velocity fields, monochromatic 
images and PV diagrams of the SINGS sample is made in this appendix. The galaxies \mbox{NGC 0925}, \mbox{NGC 2403}, \mbox{NGC 3198}, \mbox{NGC 4236}, \mbox{NGC 4321} and \mbox{NGC 6964} are part of the \BB\, survey and are extensively described in \citeauthor{2005MNRAS.360.1201H} (\citeyear{2005MNRAS.360.1201H}). The galaxies \mbox{NGC 4321}, \mbox{NGC 4536}, \mbox{NGC 4569} and \mbox{NGC 4579} are part of the sample of 30 Virgo cluster galaxies and will be studied in \citeauthor{2005astro.ph.11417C} (\citeyear{2005astro.ph.11417C}).

\textbf{NGC 628 (M74):} 
The \hI\, PA of this face-on galaxy varies greatly with distance from the center, as observed 
by \citeauthor{1992A&A...253..335K} (\citeyear{1992A&A...253..335K}). In the visible, 
this phenomenon is seen in the outer rings of the \ha\, RV map. Some \hII\, regions with velocities 
that are perpendicular to the plane of the galaxy are also visible. The \hI\, \PA\, determined by \citeauthor{1992A&A...253..335K} (\citeyear{1992A&A...253..335K}) agrees with the kinematical one (Table \ref{kinematics}), but the inclination differs greatly (6.5\Deg\, for \hI\, and 21.5\Deg\, for \ha). This may be due to the fact that \verb|rotcur| has problems dealing with galaxies whose inclination is $<$40\Deg\, (\citeauthor{1989A&A...223...47B} \citeyear{1989A&A...223...47B}, \citeauthor{2005astro.ph.11417C} \citeyear{2005astro.ph.11417C}). For the rotation curve presented in Figure \ref{figure_rotation_curves1}, the \hI\, inclination was adopted. This galaxy has also been observed in CO by \citeauthor{2001PASJ...53..757N} (\citeyear{2001PASJ...53..757N}).

\textbf{NGC 925:} 
This late type SBcd galaxy has a bright optical
and \ha\, bar and two bright patchy spiral arms beginning at the
ends of the bar. Many \hII\, regions lie along the bar. The photometrical
and kinematical data agree. The PV diagram shows non axisymmetric
motions near the center. It is well studied in \hI\, 
(\citeauthor{1998ApJ...494L..37E} \citeyear{1998ApJ...494L..37E};
\citeauthor{1998AJ....115..975P} \citeyear{1998AJ....115..975P}),
in CO (\citeauthor{2003ApJS..145..259H} \citeyear{2003ApJS..145..259H}) and
in \ha\, (\citeauthor{1982A&A...108..134M} \citeyear{1982A&A...108..134M}).
It shows strong streaming motions.

\textbf{NGC 2403:} 
This SABc galaxy shows amorphous spiral features. The \ha\, velocity maps
and the PV diagram show an almost rigid structure near the center of the galaxy. Bright \hII\, 
regions can be seen in the \ha\, monochromatic image. It is not clear whether this galaxy is barred
or not. According to  \citeauthor{1997MNRAS.292..349S} (\citeyear{1997MNRAS.292..349S}), their
Fourier harmonic analysis of the \hI\, velocity field shows that non-circular motions are not
important in this galaxy.
Moreover, \citeauthor{2000A&A...356L..49S} (\citeyear{2000A&A...356L..49S}) stress that
the thin hydrogen disk of NGC 2403 is surrounded by a vertically extended layer of \hI\, that
rotates slower than the disk. A complete modeling of the galaxy will provide more details
on its structures. \citeauthor{2001ApJ...562L..47F}  (\citeyear{2001ApJ...562L..47F}) suggest
that this anomalous \hI\, component may be similar to a class of high velocity clouds observed
in the Milky Way. In CO data, no molecular gas is detected (Helfer et al. 2003).

\textbf{NGC 2798:} 
This galaxy is interacting with its close companion, NGC 2799 on the east. 
\citeauthor{1996A&AS..120....1M} (\citeyear{1996A&AS..120....1M}) observed a difference of 125\kms\, in the velocity of the \hI\, and \hII\, components of NGC 2799. Due to this interaction, it was useless to determine the kinematical parameters of this galaxy (Table \ref{kinematics}).

\textbf{NGC 2915:}
The radial velocity map of this galaxy shows a highly distorted optical disk. This renders impossible the determination of the kinematical parameters of the galaxy. Two bright \hII\, regions are visible toward the center of the galaxy. \citeauthor{1996AJ....111.1551M} (\citeyear{1996AJ....111.1551M}) obtained the \hI\, kinematics of this blue compact 
dwarf galaxy. They observed that the optical disk of the galaxy corresponds to 
the central \hI\, bar. They think that the dark matter halo dominates at nearly all radii.

\textbf{NGC 2976:}
This peculiar dwarf late type galaxy has a nearly linear rotation curve. There is no spiral arm visible. Two strong \hII\, regions are located on each side of the galaxy. 
\citeauthor{2002A&A...389...42S} (\citeyear{2002A&A...389...42S}) observed that in \hI\, 
the rotation curve seems to flatten near the edge of the \hI\, disk. According to \citeauthor{1992AN....313....1B} (\citeyear{1992AN....313....1B}), the outer 
parts of NGC 2976 have been undisturbed for a long time and are very old (5 Gy, 
probably up to 15 Gy).

\textbf{NGC 3031 (M81):} 
The great M81 spiral galaxy has few \ha\, emission in its center, given its somewhat early type (Sab). The velocity information for radii up to $\sim$4\Min\, is 
thus difficult to extract. Farther away from the center, the rotation curve is very flat and does not show 
any decrease near the edge of the optical disk. Long slit observations performed by \citeauthor{1982A&A...106..214P} (\citeyear{1982A&A...106..214P}) show the same flattening. The bright core, as seen in infrared, contrasts greatly with its dim \ha\, counterpart. Detailed investigation of the UV, \ha\, and IR SFR indicators based on SPITZER and SINGS ancillary data have been done by \citeauthor{2004ApJS..154..215G} (\citeyear{2004ApJS..154..215G}) and suggests that the central dust is heated by stars in the bulge rather than star formation. The morphological analysis of the IR data that has been done by \citeauthor{2004ApJS..154..222W} (\citeyear{2004ApJS..154..222W}) and shows evolved stars organized in bulge and disk components, a dusty interstellar medium showing star forming regions and a clumpy profile. Still according to \citeauthor{2004ApJS..154..222W}, the flux density of the pointlike nucleus seems to have decreased by a factor of three in the past four years.

\textbf{NGC 3049:}
This Makarian galaxy harbours a ``double nucleus", as stated by \citeauthor{1995ApJS...99..461N} (\citeyear{1995ApJS...99..461N}). This feature, invisible in infrared, is easily seen in \ha\, and it does not seem to affect the galaxy's kinematics. The ``second nucleus" is most probably just a strong starburst \hII\, region. Still according to Nordgen, this galaxy does not show any trace of merging. The galaxy is totally dominated by a bar and the signal is too weak outside the bar to resolve the kinematics, rendering the extraction of kinematical parameters impossible.

\textbf{UGC 5423 (M81 dwarf B):} 
We present the first available kinematical data for this dwarf galaxy. Rotation is weak.

\textbf{NGC 3184:}
CO kinematics have been obtained by \citeauthor{2001PASJ...53..757N} (\citeyear{2001PASJ...53..757N}) for this nearly face-on galaxy. \ha\, data show a pretty flat rotation curve that nearly reaches a flat part within the optical disk.

\textbf{NGC 3198:}
This SB(rs)c galaxy has been extensively studied in \hI\, (\citeauthor{1981AJ.....86.1791B}
\citeyear{1981AJ.....86.1791B}; \citeauthor{1989A&A...223...47B} \citeyear{1989A&A...223...47B}),
FP \ha\, (\citeauthor{1991A&A...244...27C} \citeyear{1991A&A...244...27C},
Blais-Ouellette et al. 1999) and \ha\, and [\nII] long-slit spectroscopy
(\citeauthor{1998PASJ...50..427S} \citeyear{1998PASJ...50..427S}, \citeauthor{2004AJ....127.3273V} \citeyear{2004AJ....127.3273V}).
According to the PV diagram, non circular motions  near the centre can be seen. A strong
velocity gradient is also seen perpendicular to the bar major axis.

\textbf{NGC 3521:} 
The declining \hI\, rotation curve of this galaxy allowed \citeauthor{1991AJ....101.1231C} (\citeyear{1991AJ....101.1231C}) to call for the end of the ``disk-halo conspiracy". Though not visible in \ha, the rotation curve starts to decline within a radius of 22 kpc and the \ha\, data stops at 13 kpc.

\textbf{NGC 3621:}
Many strong \hII\, regions are visible in this galaxy at all galactic radii. The \ha\, data do not seem to reach the flat part of the rotation curve. This galaxy has been observed by using three fields of the ESO/La Silla \mbox{3.6-m} telescope. More kinematical data could be gathered by observing further north and south of the galaxy.

\textbf{NGC 3938:} 
The nearly face-on orientation of this galaxy allowed \citeauthor{1982A&A...105..351V} (\citeyear{1982A&A...105..351V}) to 
study it in \hI\, to search for extra-planar velocity components. An in-depth study of the \ha\, velocity map is required in order to corroborate this. 

\textbf{NGC 4236:} 
This late type SBdm galaxy is seen nearly edge-on. Its kinematical
inclination is 76\Deg. The \ha\, image shows that the \hII\, regions are distributed along the bar,
with two bright regions near the end of the bar. These features are also seen in
\hI\, (\citeauthor{1973A&A....24..411S} \citeyear{1973A&A....24..411S}). An extensive region
of solid-body rotation coincides with the bar.

\textbf{NGC 4321 (M100):} 
This grand-design spiral galaxy is located in the Virgo Cluster. It has 
been frequently mapped in the \ha\,
emission line using high-resolution FP
interferometry (\citeauthor{1990A&A...234...23A} \citeyear{1990A&A...234...23A},
\citeauthor{1990A&AS...83..211C} \citeyear{1990A&AS...83..211C}, \citeauthor{1997ApJ...479..723C} \citeyear{1997ApJ...479..723C},
\citeauthor{2000ApJ...528..219K}  \citeyear{2000ApJ...528..219K}), in the molecular CO
emission-line (\citeauthor{1990PhDT.........6C} \citeyear{1990PhDT.........6C},
\citeauthor{1995AJ....110.2075S} \citeyear{1995AJ....110.2075S}, \citeauthor{1995AJ....109.2444R}
\citeyear{1995AJ....109.2444R}, \citeauthor{1998A&A...333..864G} \citeyear{1998A&A...333..864G},
Helfer et al. 2003) and in the 21-cm \hI\, emission-line (\citeauthor{1990AJ....100..604C}
\citeyear{1990AJ....100..604C}, \citeauthor{1993ApJ...416..563K} \citeyear{1993ApJ...416..563K}).
The \hI\ disk is almost totally confined within the optical one but with a slight lopsidedness
towards the SW (\citeauthor{1993ApJ...416..563K} \citeyear{1993ApJ...416..563K}).
The \hI, CO and \ha\, velocity fields show kinematical disturbances such as streaming motions along the spiral arms and a central S-shape
distortion of the iso-velocity contours along the bar axis. The circum-nuclear region
shows the presence of an enhanced star formation region as a four-armed \ha\, ring-like structure
and a CO \& \ha\, spiral-like structure. Much more details can be found in \citeauthor{2005MNRAS.360.1201H} (\citeyear{2005MNRAS.360.1201H}) and in \citeauthor{2005astro.ph.11417C} (\citeyear{2005astro.ph.11417C}).

\textbf{NGC 4536:} 
Streaming motions along the spiral arms and a Z-shape of the velocities 
in the central parts are observed in this barred galaxy.
As in the CO data (\citeauthor{2003PASJ...55...17S}, \citeyear{2003PASJ...55...17S}), a steep velocity gradient is observed 
in the \ha\ data.

\textbf{NGC 4625:} 
This galaxy has a close companion, lying 8\Min\, away, which is only 22 kpc distant. The galaxy has a very weak rotation and harbours a lot of double profile emission lines, a sign of non-circular activity. This explains the large errors on the kinematical parameters found and the ``boiling" aspect of the \ha\, velocity field extracted. \citeauthor{1983A&AS...54...19V} (\citeyear{1983A&AS...54...19V}) studied it in \hI\, and found a neutral hydrogen disk having a diameter of 5\Min, which is $\sim$5 times larger than the observed \ha\, disk. GALEX images of this galaxy show a very extended UV disk, extending at least 2--3 times the radius of the main star-forming disk (\citeauthor{2005nowhere.1..6A} \citeyear{2005nowhere.1..6A}).

\textbf{NGC 4569 (M90):} 
This galaxy is located in the Virgo cluster. An off-plane sructure to the West of the disk of NGC 4569 has been seen through deep \ha\ imaging (\citeauthor{2001A&A...380...40T}, \citeyear{2001A&A...380...40T}) and in \hI\ data (\citeauthor{2004A&A...419...35V}, \citeyear{2004A&A...419...35V}). It is observed here as a string of \hII\, regions whose kinematics follows the rotation of the disk, although it is slightly more red-shifted than inside the disk at equal azimuth angles (\citeauthor{2005astro.ph.11417C} \citeyear{2005astro.ph.11417C}).

\textbf{NGC 4579 (M58):} 
In addition to the main large-scale spiral arms, 
 this Virgo galaxy exhibits a nuclear spiral structure (the so-called ``loop'' 
in \citeauthor{1989ApJS...71..433P}, \citeyear{1989ApJS...71..433P} and \citeauthor{1996MNRAS.281.1105G}, \citeyear{1996MNRAS.281.1105G}) 
 within which is detected a gradient of up to $\sim 500$ \kms. 
 The FP velocity field shows  that the kinematical PA of this nuclear spiral 
differs  by $\sim 90\degr$ from that of the main spiral arms. The kinematical parameters of this galaxy shown in Table \ref{kinematics} are calculated for the grand-design spiral structure (ie. outside of the nuclear spiral structure). See Figure \ref{ngc4579}. The \PA\, of the nuclear structure is $\simeq$174\Deg$\pm$11 (\citeauthor{2005astro.ph.11417C} \citeyear{2005astro.ph.11417C}).

\textbf{NGC 4725:}
This barred ringed Sab lenticular galaxy is catalogued as a double barred galaxy by \citeauthor{2004A&A...415..941E} (\citeyear{2004A&A...415..941E}) and \hI\, rich by \citeauthor{2003ApJ...585..256R} (\citeyear{2003ApJ...585..256R}). No kinematical data are available for this galaxy. The kinematics in the center of the galaxy are hard to resolve given its early type.

\textbf{NGC 5055 (M63):}
This galaxy shows several arm patterns that are well visible in \ha. It harbours a very strong velocity gradient in its center and a flat rotation curve. No bar structure is visible. It has been studied in \hI\, by \citeauthor{1981AJ.....86.1791B}  (\citeyear{1981AJ.....86.1791B}) and in CO by \citeauthor{2001PASJ...53..757N} (\citeyear{2001PASJ...53..757N}). This galaxy has also been studied in \ha\, by \citeauthor{2004A&A...420..147B} (\citeyear{2004A&A...420..147B}), which showed two velocity components in its central region. The \ha\, images presented in this paper have weaker response on the receding side of the galaxy than on the approaching side due to the wide span in the galaxy's velocities which made the receding emission fall on the wing of the interference filter.

\textbf{NGC 5194 (M51a):} 
Two emission lines are visible in the very center of the galaxy and may account for the ``dip" in the rotation curve shown in the PV diagram. The strong \hII\, regions discriminate the great spiral structure from the rest of the galaxy. A flow of \hII\, regions is seen extending towards its companion. This galaxy has been studied in \hI\, by \citeauthor{1990AJ....100..387R} (\citeyear{1990AJ....100..387R}), in both \hI\, and \ha\, (scanning FP) by \citeauthor{1991A&A...244....8T} (\citeyear{1991A&A...244....8T}) and more recently in CO by \citeauthor{1997PASJ...49..279K} (\citeyear{1997PASJ...49..279K}). NGC 5195, the galaxy's companion, has been observed in \ha\, through this study but it is 
impossible to get kinematical information as it is an early type galaxy (SB0p).

\textbf{NGC 5398:} 
This study provides the first kinematical data for this galaxy. Its peculiar radial velocity map makes it impossible to extract kinematical parameters. It has been classified as ringed galaxy by \citeauthor{1995ApJS...96...39B} (\citeyear{1995ApJS...96...39B}) and Wolf-Rayet galaxy by \citeauthor{1991ApJ...377..115C} (\citeyear{1991ApJ...377..115C}).

\textbf{NGC 5713:}
Many strong \hII\, regions are visible in this galaxy and produces a oddly looking radial velocity map. There were no kinematical data available for this galaxy prior to publishing this paper.

\textbf{IC 4710 :}
This galaxy has plenty of \hII\, regions but does not seem to harbour large-scale rotation.

\textbf{NGC 6946:}
According to \hI\, studies (\citeauthor{1990A&A...234...43C} \citeyear{1990A&A...234...43C}), the
\hI\, distribution is not symmetric but is more extended to the NE side. This feature is also
seen in the \ha\, emission map. The overall \ha\, velocity map is regular but shows some
non-circular motions near the center, confirmed by the PV diagram. It has been recently observed
in FP by \citeauthor{2004A&A...420..147B} (\citeyear{2004A&A...420..147B}) leading to the same
conclusions. Once again the wide field of \FM\, and its high sensitivity is clearly an advantage
to obtain better \ha\, velocity fields. CO data has been gathered by \citeauthor{1995ApJS...98..219Y} (\citeyear{1995ApJS...98..219Y}).

\textbf{NGC 7331:}
The receding part of the galaxy is invisible in the RV maps presented in this paper. It was first thought that this part of the galaxy was out of the interference filter, but this feature has also been observed by \citeauthor{1994A&A...282..363M} (\citeyear{1994A&A...282..363M}). An \textit{a posteriori} scan of the interference filter used showed that the galaxy should have been well centered in the filter. Deep \ha\, images taken by \citeauthor{2004ApJS..154..204R} (\citeyear{2004ApJS..154..204R}) at the KPNO 2.1 meter telescope show the same asymmetric emission pattern. However, \citeauthor{2004ApJS..154..204R} (\citeyear{2004ApJS..154..204R}) also made $\mathrm{Pa\alpha}$ observations and did not observe that asymmetry. A theory to explain this is that a ring of dust located south of the center of the galaxy is blocking the \ha\, emission from reaching us whilst letting the $\mathrm{Pa\alpha}$ through. This galaxy has been called ``Post starburst" by \citeauthor{1997ApJ...484..664T} (\citeyear{1997ApJ...484..664T}) who studied its kinematics in CO. CO data have also been gathered by \citeauthor{2001PASJ...53..757N} (\citeyear{2001PASJ...53..757N}).

\clearpage

\section[Observational data]{Observational data}
Figures were removed for the astro-ph version of this paper due to file size constraints. You can access them online at \url{http://www.astro.umontreal.ca/fantomm/sings}.

\section[Rotation curves]{Rotation curves}
Figures were removed for the astro-ph version of this paper due to file size constraints. You can access them online at \url{http://www.astro.umontreal.ca/fantomm/sings}. The rotation curves in their electronic form are available at \url{http://www.astro.umontreal.ca/fantomm/sings/rotation_curves.htm}.

\bsp

\label{lastpage}

\begin{thebibliography}{}

\bibitem[\protect\citeauthoryear{Arsenault, Roy, \&
Boulesteix}{1990}]{1990A&A...234...23A} Arsenault R., Roy J.-R., Boulesteix
J. 1990, A\&A, 234, 23

\bibitem[\protect\citeauthoryear{Begeman}{1989}]{1989A&A...223...47B} 
Begeman K.~G. 1989, A\&A, 223, 47 


\bibitem[\protect\citeauthoryear{Blais-Ouellette et 
al.}{1999}]{1999AJ....118.2123B} Blais-Ouellette S., Carignan C., Amram P., 
C{\^ o}t{\' e} S., 1999, AJ, 118, 2123 

\bibitem[\protect\citeauthoryear{Blais-Ouellette et 
al.}{2004}]{2004A&A...420..147B} Blais-Ouellette S., Amram P., Carignan C., 
Swaters R., 2004, A\&A, 420, 147

\bibitem[\protect\citeauthoryear{Boselli}{2001}]{2001ApSSS.277..401B} 
Boselli A., 2001, ApSSS, 277, 401 

\bibitem[\protect\citeauthoryear{Bosma}{1981}]{1981AJ.....86.1791B} Bosma 
A., 1981, AJ, 86, 1791 

\bibitem[\protect\citeauthoryear{Bosma et al.}{1992}]{1992ApJ...400L..21B} 
Bosma A., Byun Y., Freeman K.~C., Athanassoula E., 1992, ApJ, 400, L21 


\bibitem[\protect\citeauthoryear{Broeils \& van 
Woerden}{1994}]{1994A&AS..107..129B} Broeils A.~H., van Woerden H., 1994, 
A\&AS, 107, 129 

\bibitem[\protect\citeauthoryear{Bronkalla, Notni, \& 
Mutter}{1992}]{1992AN....313....1B} Bronkalla W., Notni P., Mutter 
A.~A.-R., 1992, AN, 313, 1 

\bibitem[\protect\citeauthoryear{del Burgo et 
al.}{2003}]{2003MNRAS.346..403D} del Burgo C., Laureijs R.~J., 
{\'A}brah{\'a}m P., Kiss C., 2003, MNRAS, 346, 403 

\bibitem[\protect\citeauthoryear{Buta}{1995}]{1995ApJS...96...39B} Buta R., 
1995, ApJS, 96, 39 

\bibitem[\protect\citeauthoryear{Canzian}{1990}]{1990PhDT.........6C}
Canzian B.~J., 1990, Ph.D.~Thesis, California Inst. of Tech., Pasadena.

\bibitem[\protect\citeauthoryear{Canzian \&
Allen}{1997}]{1997ApJ...479..723C} Canzian B., Allen R.~J. 1997, ApJ, 479,
723

\bibitem[\protect\citeauthoryear{Cappellari \& 
Copin}{2003}]{2003MNRAS.342..345C} Cappellari M., Copin Y., 2003, MNRAS, 
342, 345 

\bibitem[\protect\citeauthoryear{Carignan et
al.}{1990}]{1990A&A...234...43C} Carignan C., Charbonneau P., Boulanger F.,
Viallefond F. 1990, A\&A, 234, 43

\bibitem[\protect\citeauthoryear{Casertano \& van 
Gorkom}{1991}]{1991AJ....101.1231C} Casertano S., van Gorkom J.~H., 1991, 
AJ, 101, 1231 

\bibitem[\protect\citeauthoryear{Cayatte et
al.}{1990}]{1990AJ....100..604C} Cayatte V., van Gorkom J.~H., Balkowski
C., Kotanyi C. 1990, AJ, 100, 604

\bibitem[\protect\citeauthoryear{Cepa \&
Beckman}{1990}]{1990A&AS...83..211C} Cepa J., Beckman J.~E. 1990, A\&AS,
83, 211

\bibitem[\protect\citeauthoryear{Chemin et al.}{2006}]{2005astro.ph.11417C} 
Chemin L., et al., 2006, MNRAS accepted, arXiv:astro-ph/0511417 




\bibitem[\protect\citeauthoryear{Conti}{1991}]{1991ApJ...377..115C} Conti 
P.~S., 1991, ApJ, 377, 115 

\bibitem[\protect\citeauthoryear{Corradi et 
al.}{1991}]{1991A&A...244...27C} Corradi R.~L.~M., Boulesteix J., Bosma A., 
Amram P., Capaccioli M. 1991, A\&A, 244, 27 

\bibitem[\protect\citeauthoryear{Daigle et al.}{2006}]{2005nowhere.2..2A} Daigle, O., Carignan, C., Hernandez, O., Chemin L., Amram, P., submitted to MNRAS

\bibitem[\protect\citeauthoryear{Elmegreen, Wilcots, \& 
Pisano}{1998}]{1998ApJ...494L..37E} Elmegreen B.~G., Wilcots E., Pisano 
D.~J. 1998, ApJ, 494, L37 

\bibitem[\protect\citeauthoryear{Erwin}{2004}]{2004A&A...415..941E} Erwin 
P., 2004, A\&A, 415, 941 

\bibitem[\protect\citeauthoryear{Fraternali et
al.}{2001}]{2001ApJ...562L..47F} Fraternali F., Oosterloo T., Sancisi R.,
van Moorsel G. 2001, ApJ, 562, L47

\bibitem[\protect\citeauthoryear{Gach et al.}{2002}]{2002PASP..114.1043G}  Gach J.-L., Hernandez, O.; Boulesteix, J.; Amram, P.; Boissin, O.; Carignan, C.; Garrido, O.; Marcelin, M.; \"Ostlin, G.; Plana, H.; Rampazzo, R., 2002, PASP, 114, 1043 

\bibitem[\protect\citeauthoryear{Garcia-Burillo et
al.}{1998}]{1998A&A...333..864G} Garcia-Burillo S., Sempere M.~J., Combes
F., Neri R. 1998, A\&A, 333, 864

\bibitem[\protect\citeauthoryear{Gil de Paz et al.}{2005}]{2005nowhere.1..6A} Gil de Paz et al., submitted to ApJ

\bibitem[\protect\citeauthoryear{Goad \& 
Roberts}{1981}]{1981ApJ...250...79G} Goad J.~W., Roberts M.~S., 1981, ApJ, 
250, 79 

\bibitem[\protect\citeauthoryear{Gonzalez Delgado \& 
Perez}{1996}]{1996MNRAS.281.1105G} Gonzalez Delgado R.~M., Perez E., 1996, 
MNRAS, 281, 1105 

\bibitem[\protect\citeauthoryear{Gordon et al.}{2004}]{2004ApJS..154..215G} 
Gordon K.~D., et al., 2004, ApJS, 154, 215 

\bibitem[Helfer et al.(2003)]{2003ApJS..145..259H} Helfer, T.~T., Thornley, 
M.~D., Regan, M.~W., Wong, T., Sheth, K., Vogel, S.~N., Blitz, L., \& Bock, 
D.~C.-J.\ 2003, ApJS, 145, 259 

\bibitem[\protect\citeauthoryear{Hernandez et 
al.}{2003}]{2003SPIE.4841.1472H} Hernandez O., Gach J., Carignan C., 
Boulesteix J., 2003, SPIE, 4841, 1472 

\bibitem[\protect\citeauthoryear{Hernandez et 
al.}{2005}]{2005MNRAS.360.1201H} Hernandez O., Carignan C., Amram P., 
Chemin L., Daigle O., 2005, MNRAS, 360, 1201 


\bibitem[\protect\citeauthoryear{Hernandez et 
al.}{2005}]{2005ApJ...632..253H} Hernandez O., Wozniak H., Carignan C., 
Amram P., Chemin L., Daigle O., 2005, ApJ, 632, 253 


\bibitem[\protect\citeauthoryear{Huchtmeier}{1975}]{1975A&A....45..259H} 
Huchtmeier W.~K., 1975, A\&A, 45, 259 

\bibitem[\protect\citeauthoryear{Kamphuis \& 
Briggs}{1992}]{1992A&A...253..335K} Kamphuis J., Briggs F., 1992, A\&A, 
253, 335 

\bibitem[\protect\citeauthoryear{Kennicutt}{1989}]{1989ApJ...344..685K} 
Kennicutt R.~C., 1989, ApJ, 344, 685


\bibitem[\protect\citeauthoryear{Kennicutt et 
al.}{2003}]{2003PASP..115..928K} Kennicutt R.~C., et al., 2003, PASP, 115, 
928 

\bibitem[\protect\citeauthoryear{Knapen et al.}{1993}]{1993ApJ...416..563K}
Knapen J.~H., Cepa J., Beckman J.~E., Soledad del Rio M., Pedlar A. 1993,
ApJ, 416, 563

\bibitem[\protect\citeauthoryear{Knapen et al.}{2000}]{2000ApJ...528..219K}
Knapen J.~H., Shlosman I., Heller C.~H., Rand R.~J., Beckman J.~E., Rozas
M. 2000, ApJ, 528, 219

\bibitem[\protect\citeauthoryear{Kuno \& Nakai}{1997}]{1997PASJ...49..279K} 
Kuno N., Nakai N., 1997, PASJ, 49, 279 

\bibitem[\protect\citeauthoryear{Marcelin, Boulesteix, \& 
Courtes}{1982}]{1982A&A...108..134M} Marcelin M., Boulesteix J., Courtes 
G. 1982, A\&A, 108, 134 

\bibitem[\protect\citeauthoryear{Marcelin et 
al.}{1994}]{1994A&A...282..363M} Marcelin M., Petrosian A.~R., Amram P., 
Boulesteix J., 1994, A\&A, 282, 363 

\bibitem[\protect\citeauthoryear{Marquez \& 
Moles}{1996}]{1996A&AS..120....1M} Marquez I., Moles M., 1996, A\&AS, 120, 
1 

\bibitem[\protect\citeauthoryear{Meurer et al.}{1996}]{1996AJ....111.1551M} 
Meurer G.~R., Carignan C., Beaulieu S.~F., Freeman K.~C., 1996, AJ, 111, 
1551 

\bibitem[\protect\citeauthoryear{Nishiyama, Nakai, \& 
Kuno}{2001}]{2001PASJ...53..757N} Nishiyama K., Nakai N., Kuno N., 2001, 
PASJ, 53, 757 

\bibitem[\protect\citeauthoryear{Nordgren et 
al.}{1995}]{1995ApJS...99..461N} Nordgren T.~E., Helou G., Chengalur J.~N., 
Terzian Y., Khachikian E., 1995, ApJS, 99, 461 

\bibitem[\protect\citeauthoryear{Pellet \& 
Simien}{1982}]{1982A&A...106..214P} Pellet A., Simien F., 1982, A\&A, 106, 
214 

\bibitem[\protect\citeauthoryear{Pisano, Wilcots, \& 
Elmegreen}{1998}]{1998AJ....115..975P} Pisano D.~J., Wilcots E.~M., 
Elmegreen B.~G. 1998, AJ, 115, 975 

\bibitem[\protect\citeauthoryear{Pogge}{1989}]{1989ApJS...71..433P} Pogge 
R.~W., 1989, ApJS, 71, 433 

\bibitem[\protect\citeauthoryear{Rand}{1995}]{1995AJ....109.2444R} Rand
R.~J. 1995, AJ, 109, 2444

\bibitem[\protect\citeauthoryear{Regan et al.}{2004}]{2004ApJS..154..204R} 
Regan M.~W., et al., 2004, ApJS, 154, 204 

\bibitem[\protect\citeauthoryear{Rosenberg \& 
Schneider}{2003}]{2003ApJ...585..256R} Rosenberg J.~L., Schneider S.~E., 
2003, ApJ, 585, 256 .

\bibitem[\protect\citeauthoryear{Rots et al.}{1990}]{1990AJ....100..387R} 
Rots A.~H., Crane P.~C., Bosma A., Athanassoula E., van der Hulst J.~M., 
1990, AJ, 100, 387 

\bibitem[\protect\citeauthoryear{Sakamoto et
al.}{1995}]{1995AJ....110.2075S} Sakamoto K., Okumura S., Minezaki T.,
Kobayashi Y., Wada K. 1995, AJ, 110, 2075

\bibitem[\protect\citeauthoryear{Sandage}{1986}]{1986ARA&A..24..421S} 
Sandage A., 1986, ARA\&A, 24, 421 

\bibitem[\protect\citeauthoryear{Schaap, Sancisi, \&
Swaters}{2000}]{2000A&A...356L..49S} Schaap W.~E., Sancisi R., Swaters
R.~A. 2000, A\&A, 356, L49

\bibitem[\protect\citeauthoryear{Schoenmakers, Franx, \& de Zeeuw}{1997}]{1997MNRAS.292..349S} Schoenmakers R.~H.~M., Franx M., de Zeeuw P.~T. 1997, MNRAS, 292, 349

\bibitem[\protect\citeauthoryear{Shostak}{1973}]{1973A&A....24..411S} 
Shostak G.~S. 1973, A\&A, 24, 411 

\bibitem[\protect\citeauthoryear{Sofue et al.}{1998}]{1998PASJ...50..427S} 
Sofue Y., Tomita A., Tutui Y., Honma M., Takeda Y. 1998, PASJ, 50, 427 

\bibitem[\protect\citeauthoryear{Sofue et al.}{2003}]{2003PASJ...55...17S} 
Sofue Y., Koda J., Nakanishi H., Onodera S., Kohno K., Tomita A., Okumura 
S.~K., 2003, PASJ, 55, 17 

\bibitem[\protect\citeauthoryear{Stil \& 
Israel}{2002}]{2002A&A...389...42S} Stil J.~M., Israel F.~P., 2002, A\&A, 
389, 42 

\bibitem[\protect\citeauthoryear{Tremaine \& 
Weinberg}{1984}]{1984ApJ...282L...5T} Tremaine S., Weinberg M.~D., 1984, 
ApJ, 282, L5 

\bibitem[\protect\citeauthoryear{Tosaki \& 
Shioya}{1997}]{1997ApJ...484..664T} Tosaki T., Shioya Y., 1997, ApJ, 484, 
664 

\bibitem[\protect\citeauthoryear{van der Kruit \& 
Shostak}{1982}]{1982A&A...105..351V} van der Kruit P.~C., Shostak G.~S., 
1982, A\&A, 105, 351 

\bibitem[\protect\citeauthoryear{Vogt et al.}{2004}]{2004AJ....127.3273V}
Vogt N.~P., Haynes M.~P., Herter T., Giovanelli R. 2004, AJ, 127, 3273

\bibitem[\protect\citeauthoryear{Tilanus \& 
Allen}{1991}]{1991A&A...244....8T} Tilanus R.~P.~J., Allen R.~J., 1991, 
A\&A, 244, 8 

\bibitem[\protect\citeauthoryear{Tsch{\" o}ke et 
al.}{2001}]{2001A&A...380...40T} Tsch{\" o}ke D., Bomans D.~J., Hensler G., 
Junkes N., 2001, A\&A, 380, 40 

\bibitem[\protect\citeauthoryear{van Moorsel}{1983}]{1983A&AS...54...19V} 
van Moorsel G.~A., 1983, A\&AS, 54, 19 

\bibitem[\protect\citeauthoryear{Vollmer et 
al.}{2004}]{2004A&A...419...35V} Vollmer B., Balkowski C., Cayatte V., van 
Driel W., Huchtmeier W., 2004, A\&A, 419, 35 

\bibitem[\protect\citeauthoryear{Wevers, van der Kruit, \& 
Allen}{1986}]{1986A&AS...66..505W} Wevers B.~M.~H.~R., van der Kruit P.~C., 
Allen R.~J., 1986, A\&AS, 66, 505 

\bibitem[\protect\citeauthoryear{Willner et 
al.}{2004}]{2004ApJS..154..222W} Willner S.~P., et al., 2004, ApJS, 154, 
222 

\bibitem[\protect\citeauthoryear{Young et al.}{1995}]{1995ApJS...98..219Y} 
Young J.~S., et al., 1995, ApJS, 98, 219 

\end{thebibliography}
\end{document}